\documentclass[reprint,prl,twocolumn,showpacs,superscriptaddress,aps,longbibliography]{revtex4-2}

\usepackage[colorlinks=true,citecolor=blue]{hyperref} 
\usepackage{graphicx}
\usepackage{bm}
\usepackage{amsmath}
\usepackage{amssymb}
\usepackage{comment}
\usepackage{braket}
\usepackage{lipsum}
\usepackage{chemformula}
\usepackage{booktabs}
\newcommand{\dIdV}{d$I$/d$V$}
\DeclareGraphicsExtensions{.png,.jpg,.eps}
\usepackage{xcolor}
\usepackage[utf8]{inputenc}

\newcommand{\compound}[1]{\textbf{#1}}
\newcommand{\nmrsolv}[1]{\text{#1}}
\usepackage{float}

\newcounter{suppsection}
\renewcommand{\thesuppsection}{S\Roman{suppsection}}

\newcommand{\suppsection}[1]{%
  \refstepcounter{suppsection}%
  \section*{\thesuppsection.\ #1}%
}

\begin{document}
\title{Learning Inhomogeneous Heisenberg Hamiltonians in Nanographene Spin Chains}

\author{Greta Lupi}
\email{greta.lupi@aalto.fi}
\affiliation{Department of Applied Physics, Aalto University, 02150 Espoo, Finland}

\author{Saketh Ravuri}
\affiliation{nanotech@surfaces Laboratory, Empa---Swiss Federal Laboratories for Materials Science and Technology, 8600 Dübendorf, Switzerland}

\author{Chenxiao Zhao}
\affiliation{nanotech@surfaces Laboratory, Empa---Swiss Federal Laboratories for Materials Science and Technology, 8600 Dübendorf, Switzerland}
\affiliation{Quantum Science Center of Guangdong-Hong Kong-Macao Greater Bay Area, Shenzhen 518045, China}

\author{Weidan Zhang}
\affiliation{Max Planck Institute of Microstructure Physics, Weinberg 2, 06120 Halle, Germany}
\affiliation{Center for Advancing Electronics Dresden (cfaed) \& Faculty of Chemistry and Food Chemistry, Technische Universität Dresden, 01062 Dresden, Germany}

\author{Cesare Roncaglia}
\affiliation{nanotech@surfaces Laboratory, Empa---Swiss Federal Laboratories for Materials Science and Technology, 8600 Dübendorf, Switzerland}

\author{Renxiang Liu}
\affiliation{Max Planck Institute of Microstructure Physics, Weinberg 2, 06120 Halle, Germany}
\affiliation{Center for Advancing Electronics Dresden (cfaed) \& Faculty of Chemistry and Food Chemistry, Technische Universität Dresden, 01062 Dresden, Germany}

\author{Xinliang Feng}
\affiliation{Max Planck Institute of Microstructure Physics, Weinberg 2, 06120 Halle, Germany}
\affiliation{Center for Advancing Electronics Dresden (cfaed) \& Faculty of Chemistry and Food Chemistry, Technische Universität Dresden, 01062 Dresden, Germany}

\author{Daniele Passerone}
\affiliation{nanotech@surfaces Laboratory, Empa---Swiss Federal Laboratories for Materials Science and Technology, 8600 Dübendorf, Switzerland}

\author{Pascal Ruffieux}
\affiliation{nanotech@surfaces Laboratory, Empa---Swiss Federal Laboratories for Materials Science and Technology, 8600 Dübendorf, Switzerland}

\author{Roman Fasel}
\affiliation{nanotech@surfaces Laboratory, Empa---Swiss Federal Laboratories for Materials Science and Technology, 8600 Dübendorf, Switzerland}
\affiliation{Department of Chemistry, Biochemistry and Pharmaceutical Sciences, University of Bern, 3012 Bern, Switzerland}

\author{Jose L. Lado}
\affiliation{Department of Applied Physics, Aalto University, 02150 Espoo, Finland}

\author{Gonçalo Catarina}
\email{goncalo.catarina@empa.ch}
\affiliation{nanotech@surfaces Laboratory, Empa---Swiss Federal Laboratories for Materials Science and Technology, 8600 Dübendorf, Switzerland}

\date{\today}


\begin{abstract}
Inferring microscopic Hamiltonians from experimental data is a central challenge in quantum materials and quantum simulation. In low-dimensional spin systems, exchange interactions are often assumed to be spatially uniform, despite structural and environmental inhomogeneities that can locally modify the coupling.
Here, we leverage a local, length-independent machine learning methodology to reconstruct spatially modulated exchange interactions directly from inelastic scanning tunneling spectroscopy maps.
We demonstrate this approach with nanographene spin chains, identifying both near-uniform and inhomogeneous regimes across the synthesized magnets. 
The reconstructed models quantitatively reproduce the experimental spectra and recover the correct scaling of the excitation gap with system size.
Our results establish a general strategy to bridge local spectroscopic measurements with effective many-body Hamiltonians.
\end{abstract}

\maketitle


\textit{Introduction:} Heisenberg spin chains represent a paradigmatic model of strongly correlated quantum systems, capturing essential aspects of quantum magnetism and serving as a benchmark for quantum simulators. Recent advances in the synthesis of atomically precise nanographene structures~\cite{deOteyza_2022,Zhang2024} have enabled the realization of finite-length spin chains, where localized spin moments arise from engineered $\pi$-electron states associated with radical sites, which can be probed directly by scanning tunneling microscopy (STM) through their spectroscopic signatures~\cite{Mishra2021,Zhao2024tunable,Su2025,Kewei2025,Fu2025,Yuan2025,Zhao2025}.
The low-energy excitation spectrum is typically interpreted within an isotropic Heisenberg model with a uniform nearest-neighbor exchange interaction, which enables to qualitatively capture a wide variety
of experimental observations. However, real nanographene spin chains are not perfectly ideal: variations in molecular geometry and bond connections, local distortions, and the influence of the substrate can locally modify the electronic environment~\cite{Vegliante2024,Catarina2024,Krane2023,Zhao2024}. However, directly accessing such bond-dependent interactions from experimental data remains a highly non-trivial task, as the spectroscopic response depends non-locally on multiple, highly correlated parameters. 

\begin{figure}
    \centering
    \includegraphics[width=\linewidth]{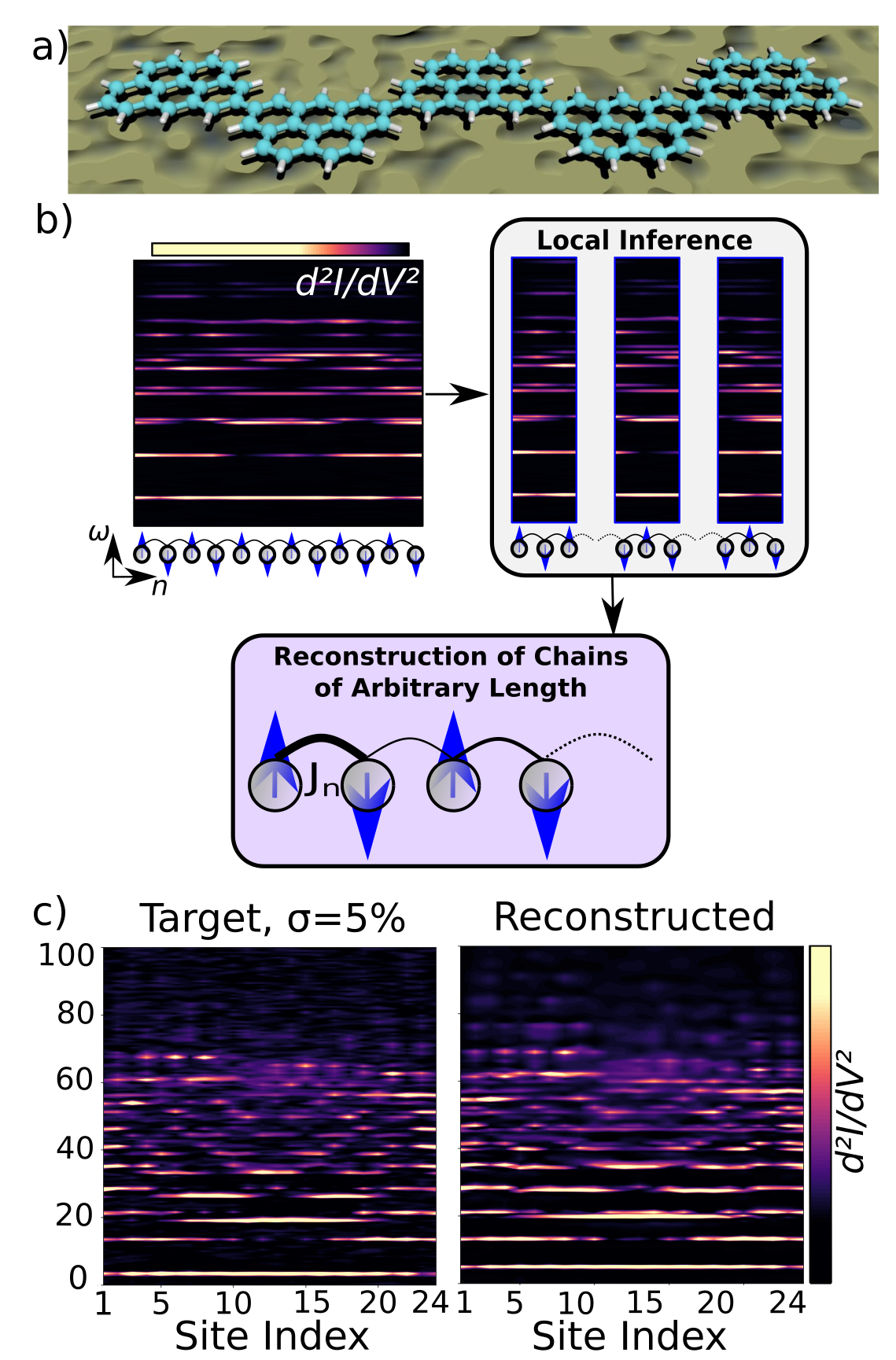}
    \caption{
    (a) Illustration of an olympicene spin chain on Au(111).
    (b) Schematic of the Hamiltonian learning framework, where spatially resolved STM spectroscopy is used to reconstruct bond-dependent exchange couplings $J_n$ via a local inference scheme. 
    (c) Target and reconstructed synthetic $d^2I/dV^2$ maps are shown in the presence of input noise with amplitude $\sigma = 5\%$ of the $dI/dV$ signal.
    }
    \label{fig:fig1}
\end{figure}

Machine learning techniques have
risen as a flexible strategy to deal with problems that are
challenging for standard methodologies
in condensed matter,
including many-body solvers~\cite{Carleo2017}, phase classification~\cite{vanNieuwenburg2017,Carrasquilla2017},
autonomous device tuning~\cite{PhysRevApplied.13.054019,2024arXiv240504596V},
autonomous single-atom manipulation~\cite{Chen2022,Wu2024},
and Hamiltonian learning~\cite{PhysRevA.105.023302,Lupi2025,Karjalainen2023,Karjalainen2025,vandriel2024,Lupi2026MolecularHL,Hernandes2026,Nigmatulin2026,Wang2017,Koch2022,Koch2023,Valenti2022,Che2021,Simard2025}.
Hamiltonian learning tackles the problem
of inferring effective microscopic models directly from experimental observables~\cite{Barey2019,Gebhart2023},
and it has been demonstrated in applications 
including quantum devices~\cite{vandriel2024}, 
correlated electron materials~\cite{Sobral2023}, and scanning probe experiments~\cite{Koch2025,Lupi2026MolecularHL}. This approach provides a powerful route to tackle inverse problems that are otherwise inaccessible through conventional fitting methods.

In this work, we demonstrate a machine learning methodology to infer spatially varying exchange interactions directly from STM spectroscopy. By exploiting a local, length-independent inference scheme~\cite{Koch2025}, we reconstruct the exchange coupling at the bond level and apply the method to experimental nanographene spin-$1/2$ chains composed of olympicene molecules (Fig.~\ref{fig:fig1}a). Our results reveal that these systems can naturally realize both near-uniform and inhomogeneous regimes. In both cases, the reconstructed model quantitatively reproduces the experimental spectroscopic maps, and, as a key outcome, also recovers the correct scaling of the excitation gap with chain length, resolving the apparent discrepancies observed within a uniform description~\cite{Zhao2025}.


\textit{Learning spatially resolved exchange interactions:} To account for inhomogeneity in realistic olympicene spin chains, we relax the assumption of a spatially uniform exchange interaction. Accordingly, we consider a spin-$1/2$ Heisenberg model in which the exchange coupling becomes bond-dependent,
\begin{equation}
H = \sum_n J_n \, \mathbf{S}_n \cdot \mathbf{S}_{n+1}.
\label{eq:inhomogeneous_heisenberg}
\end{equation}

The connection to experimental observables is established through the local dynamical spin response, defined as
\begin{equation}
A_n(\omega) = \sum_{\alpha=x,y,z}\langle \Omega | {S_n^\alpha} \, \delta(\omega - H + E_{\Omega}) \, S_n^\alpha | \Omega \rangle,
\end{equation}
where $|\Omega\rangle$ is the ground state with energy $E_{\Omega}$. This quantity captures the local excitation spectrum. In inelastic scanning tunneling spectroscopy (STS), the measured $d^2I/dV^2$ signal is proportional to this dynamical correlator, 
providing a direct experimental probe of spin excitations~\cite{Heinrich2004,Cyrus2006,Ternes_2015}.

To access the spatial modulation of the exchange interaction, we employ a data driven approach that extracts local couplings directly from STS maps, in the spirit of recent Hamiltonian learning approaches~\cite{Barey2019,Gebhart2023,Karjalainen2023,Lupi2025,Lupi2026MolecularHL,Koch2025}.
Starting from spatially resolved \dIdV maps, we perform a local analysis based on sliding windows that span three adjacent sites (Fig.~\ref{fig:fig1}b). Within each window, the measured spectrum is primarily sensitive to the exchange couplings of the corresponding bonds. By training a supervised algorithm on experimental-like data generated from Heisenberg chains with varying $J_n$, we learn a mapping from local spectra to the underlying exchange parameters~\footnote{The details of dataset generation and preprocessing are provided in the Supplemental Information (SI)}.
This procedure yields bond-resolved estimates of the exchange interaction, $J_n$, across the entire chain. The couplings from overlapping windows are subsequently combined to obtain a consistent spatial profile of $J_n$. Importantly, since the inference is performed locally, the method is intrinsically independent of the total chain length, allowing the same trained algorithm to be applied to chains of arbitrary length. The only requirement is a minimal local environment of three adjacent sites, making it suitable to chains with length $L>2$.

The validity of this approach is first evaluated on synthetic theoretical data, where the method accurately recovers randomly generated inhomogeneous exchange profiles within the training range [35,45]~meV, while remaining robust in the presence of noise and spectral broadening~\footnote{Systematic benchmarks of the trained algorithm in the SI}. These tests confirm that local spectroscopic features provide sufficient information to reconstruct the exchange interaction at the single site level. In Fig.~\ref{fig:fig1}c, we show a representative benchmark for a chain of length $L=24$, where the reconstructed spectroscopic map accurately reproduces the target in the presence of noise with amplitude corresponding to $\sigma = 5\%$ of the $dI/dV$ signal.

\begin{figure}
    \centering
    \includegraphics[width=\linewidth]{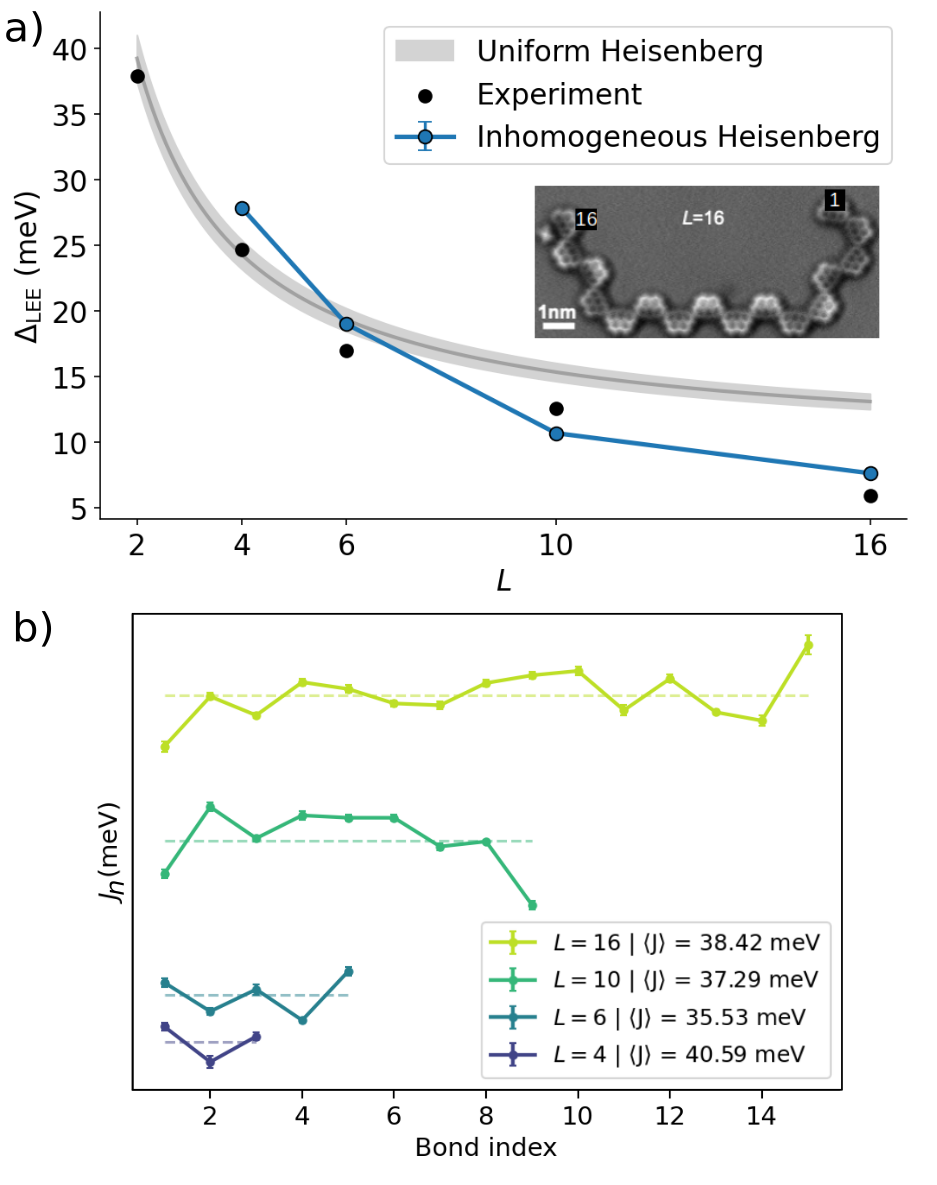}
    \caption{
    \textbf{Inhomogeneous regime (Chain IH)}. 
    (a) Lowest energy spin excitation $\Delta_{LEE}$ as a function of chain length $L$. Black circles denote measurements performed with $I_{set}=500~$pA and $V_{mod}=2$~mV. The shaded region corresponds to the range of gaps obtained with a uniform Heisenberg model using the minimum and maximum dimer exchange couplings measured across the full experimental dataset of Ref.~\cite{Zhao2025}. Blue circles denote the gaps predicted from Hamiltonian learning. Inset: nc-AFM image of Chain IH.
    (b) Bond-resolved exchange couplings $J_n$ learned from STM spectroscopy measurements, revealing significant spatial variations along the chain. 
}
    \label{fig:fig2}
\end{figure}

\textit{Application to olympicene spin chains:} We apply our framework to olympicene spin chains realized on Au(111). The data analyzed here combine previously reported measurements from Ref.~\cite{Zhao2025} and new experimental data obtained on chains synthesized as follows. Olympicene precursor molecules~\footnote{Further details on the synthesis are provided in the SI.} were deposited onto a clean Au(111) surface and annealed at $\sim 250$ °C for 5 min for debromination, followed by annealing at $\sim 350$ °C for 7 min to induce cyclodehydrogenation. The resulting chains were hydrogen-passivated and subsequently reactivated via tip-induced dehydrogenation~\cite{Zhao2024}. 

Different effective chain lengths $L$ are realized within the same nanographene structure by sequential activation of spin sites, rather than by fabricating distinct chains~\cite{Zhao2025}. After each activation step, spatially resolved spectroscopy is performed, enabling a controlled investigation of the evolution of spin excitations with $L$~\footnote{Further experimental details are provided in the SI.}.
In the following, we consider two representative cases: a strongly inhomogeneous chain (Chain IH), taken from Ref.~\cite{Zhao2025}, and a near-uniform chain (Chain H) obtained from the newly synthesized structures described above.



We begin with Chain IH. As shown in Fig.~\ref{fig:fig2}a, the experimental excitation gap $\Delta_{LEE}$ as a function of chain length $L$ cannot be quantitatively reproduced by a model with a constant exchange coupling $J$, which systematically overestimates the gap. This discrepancy does not indicate a breakdown of the Heisenberg framework, but rather a limitation of the assumption of spatially uniform interactions, as the experimental gap scaling is recovered by allowing for bond-dependent couplings.

We use our machine learning algorithms to reconstruct the exchange interaction at the bond level, $J_n$, using the local spectroscopy measurements. The resulting spatial profile exhibits pronounced variations along the chain (Fig.~\ref{fig:fig2}b), demonstrating that this system cannot be well-described by a single global exchange parameter.
As the chain length increases, the reconstructed couplings in the previously activated portion of the chain remain qualitatively consistent. This suggests that the inferred exchange profile reflects an underlying physical structure rather than random fluctuations.
Importantly, the gap for the different chains now
has reduced the mean absolute error from $\sim 2.6$~meV in the uniform model to $\sim 1.8$~meV
in the learned non-uniform one.
Based on the reconstructed couplings, we compute the corresponding excitation spectrum and spectroscopic maps. The calculated \dIdV maps are in quantitative agreement with experiment, accurately reproducing both the position and intensity of the excitation features \footnote{See Supplementary Information for a detailed comparison, including reconstructed maps and quantitative error analysis}.
These results establish that Chain IH is intrinsically inhomogeneous, and that spatial variations of the exchange interaction play a key role in determining its low-energy properties.

\begin{figure}
    \centering
    \includegraphics[width=\linewidth]{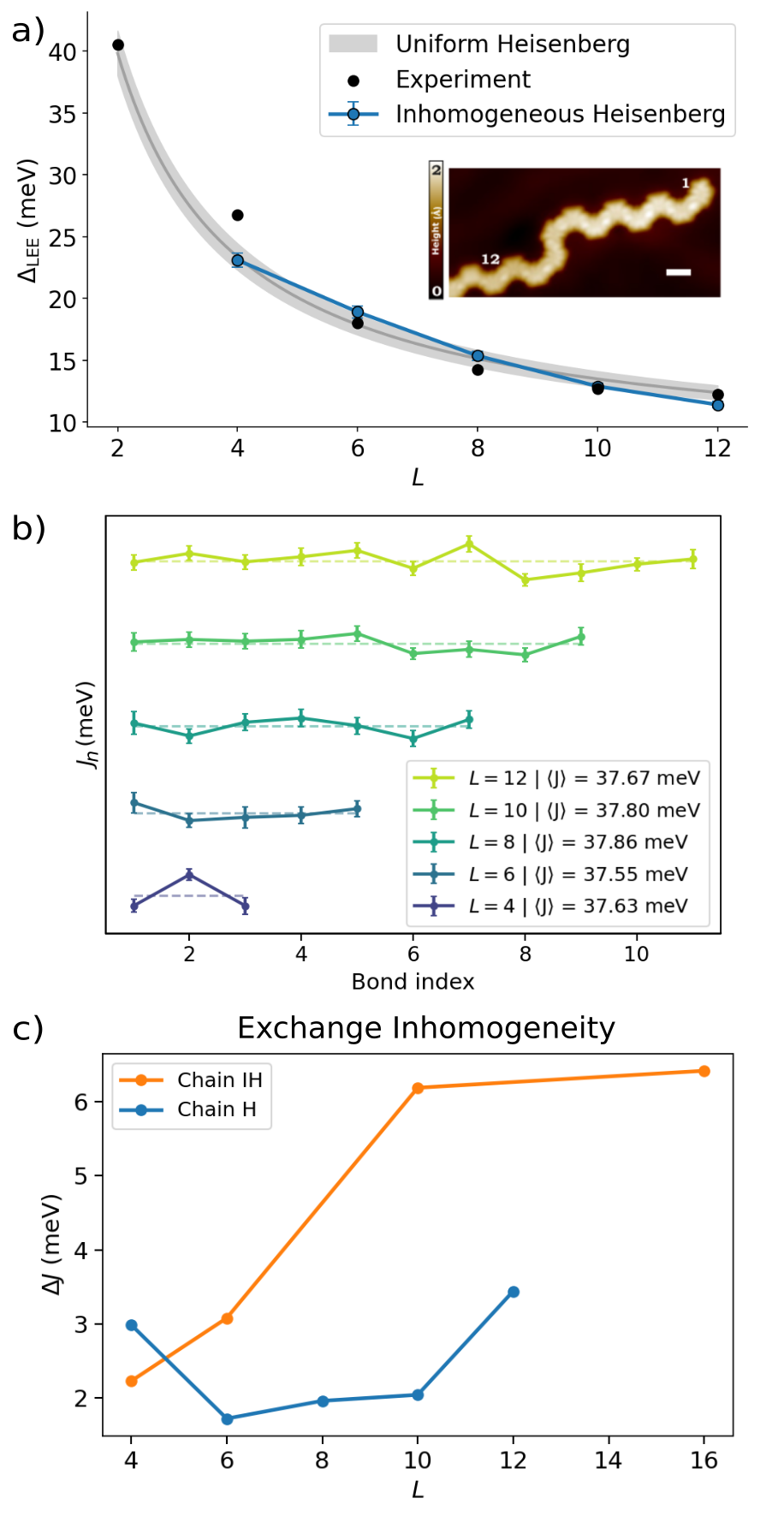}
    \caption{
    \textbf{Near-uniform regime (Chain H).} 
    (a) Spin excitation gap $\Delta_{LEE}$ as a function of chain length $L$, well described by a uniform Heisenberg model. Spectra parameters: $I_{set}=1$~nA and $V_{mod}=1$~mV. Inset: STM image of Chain H, obtained with $V_{bias}=-100$~mV and $I_{set}= 100 $~pA. Scale bar is 1~nm. 
    (b) Bond-resolved exchange couplings $J_n$, revealing small spatial variations. 
    (c) Comparison of the exchange inhomogeneity $\Delta J$ for Chains IH and H.
}
    \label{fig:fig3}
\end{figure}


We next consider Chain H. In contrast to Chain IH, the experimental excitation gap is well reproduced by a uniform Heisenberg model (Fig.~\ref{fig:fig3}a), suggesting that spatial variations of the exchange interaction are comparatively small in this case.
Applying the same reconstruction procedure, we extract the bond-resolved exchange couplings $J_n$ along the chain. The resulting profile is nearly uniform (Fig.~\ref{fig:fig3}b). This is further quantified by the  exchange inhomogeneity $\Delta J = J_{\text{max}}-J_{\text{min}}$, which remains below $\sim 3.5$~meV, in contrast to Chain IH where it exceeds $6$~meV for longer chain lengths (Fig.~\ref{fig:fig3}c). The small exchange inhomogeneity found in Chain H is comparable to typical experimental uncertainties in the extraction of exchange couplings, indicating that this chain lies close to the homogeneous limit.
Despite this near-uniform behavior, the inhomogeneous reconstruction remains fully consistent with the experimental data and reproduces the measured gap with a mean absolute error of $\sim 1.1$ meV. 

\begin{figure}
    \centering
    \includegraphics[width=
\linewidth]{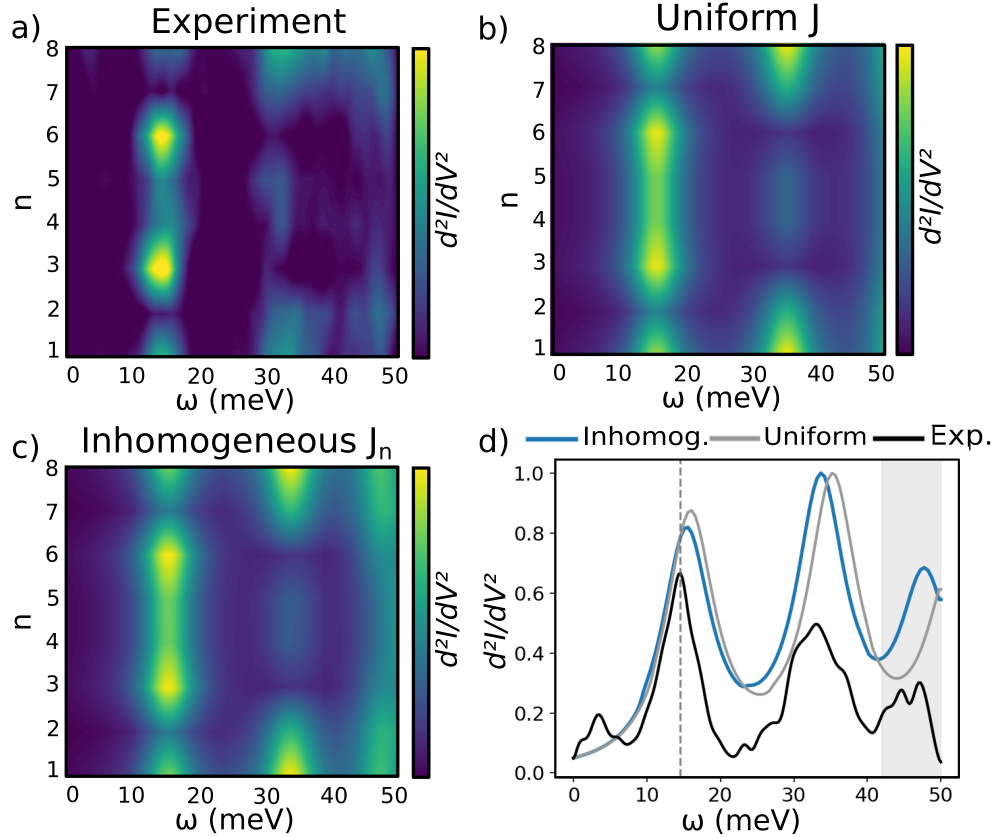}
    \caption{
    (a) Experimental d$^2I$/d$V^2$ map for Chain H with $L=8$. 
    (b) Calculated d$^2I$/d$V^2$ map assuming a uniform exchange coupling $J = 40.5$~meV, corresponding to the experimental value extracted from the dimer ($L=2$) of Chain H. 
    (c) Reconstructed d$^2I$/d$V^2$ map using the learned inhomogeneous couplings $J_n$. 
    (d) Local $d^2I/dV^2$ spectrum at site $n=1$. The dashed line indicates the experimentally measured lowest-energy excitation. The higher-energy excitation is accurately reproduced only by the inhomogeneous model.
}
    \label{fig:fig4}
\end{figure}

An interesting feature is that, even in the nearly uniform limit, disorder has an impact on the
energy excitations above the gap. This is shown in Fig.~\ref{fig:fig4}, which compares the experimental spectroscopy (panel a) with simulations for the homogeneous (panel b) and inhomogeneous (panel c) models.
The low energy peak
at $\sim15$~meV 
determines the gap $\Delta_{LEE}$, and is used to fix the exchange coupling in the uniform model. The third peak at $\sim50$~meV,
however, shows a sizable shift
between the experiment and the uniform simulation.
In contrast, the non-uniform model extracted by our machine learning algorithm reproduces this peak more accurately (Fig.~\ref{fig:fig4}d).
These results demonstrate that, even in this nearly uniform case where disorder has a minor
effect on the gap, it can remarkably affect higher energy excitations, as captured by the Hamiltonian learning algorithm.

\textit{Discussion:} The emergence of spatially varying exchange interactions in nanographene spin chains can originate from several microscopic mechanisms. A primary contribution is the interaction with the metallic substrate. Since the exchange coupling is mediated by the $\pi$-electron states, local changes in hybridization, adsorption geometry, or electrostatic environment can modify the effective overlap between neighboring spin centers. Even for nominally identical chains, the Au(111) surface contains atomic-scale variations that can perturb the local spin state and exchange pathways~\cite{Zhao2024}.
The chain geometry provides a second possible source of variation. The comparison between Chain IH and Chain H suggests that more bent or irregular configurations seem to remain closer to the homogeneous limit. A possible interpretation is that bends and local distortions help redistribute substrate-induced perturbations over several bonds, reducing the impact of a single local environment on one specific exchange coupling. 
Additional contributions may stem from activation processes, as the neighboring passivated sites adopt $sp^3$ hybridization, which drives the structure to deviate from planarity.
Overall, these results suggest that exchange inhomogeneity is not an artifact of the reconstruction procedure, but a physical feature of real nanographene spin chains on Au(111). The ability to resolve $J_n$ at the bond level therefore provides a direct way to connect atomic-scale structure, substrate environment, and the emergent excitation spectrum.


\textit{Conclusion:} In this work, we demonstrated
a machine learning strategy that allows reconstructing bond-resolved exchange couplings directly from spatially-resolved STM spectroscopy.
We applied our method to olympicene spin chains on Au(111), showing that spatial variations of the exchange interaction are essential for a quantitative description. The resulting inhomogeneous Hamiltonian consistently reproduces the full experimental spectroscopic response and, as a direct consequence, recovers the correct scaling of the excitation gap.
More broadly, our results show that spatial inhomogeneity, intrinsically expected in realistic systems, can have a measurable impact on their emergent properties, highlighting the importance of accounting for such effects when designing and interpreting molecular quantum simulators.
Our machine learning method enables a direct link between atomic-scale structure and effective many-body descriptions, opening new avenues for Hamiltonian learning and for the quantitative characterization and control of engineered quantum systems.

\textbf{Data Availability:} 
Both the source code and the datasets are publicly available on GitHub~\cite{lupi2026repo}.

\textbf{Acknowledgments:}
We acknowledge financial support from the European Research Council ERC-2024-CoG ULTRATWISTROICS (No.~101170477), the Finnish Centre of Excellence in Quantum Materials QMAT (No. 374166), the Finnish Quantum Flagship (project No.~358877), the Werner Siemens Foundation (CarboQuant), and the Empa Young Scientist Fellowship. We acknowledge the computational resources provided by the Aalto Science-IT project. 

\bibliography{ref}

\clearpage

\begin{center}
{\Large\bfseries\scshape Supplemental Material}
\end{center}
\vspace{0.5cm}

\suppsection{Synthesis and spectroscopy details}

This section provides the synthesis and measurement details for Chain H, while the synthesis and spectroscopy of Chain IH are reported in Ref.~\cite{Zhao2025}.

\begin{figure}[h]
\centering
\includegraphics[width=\linewidth]{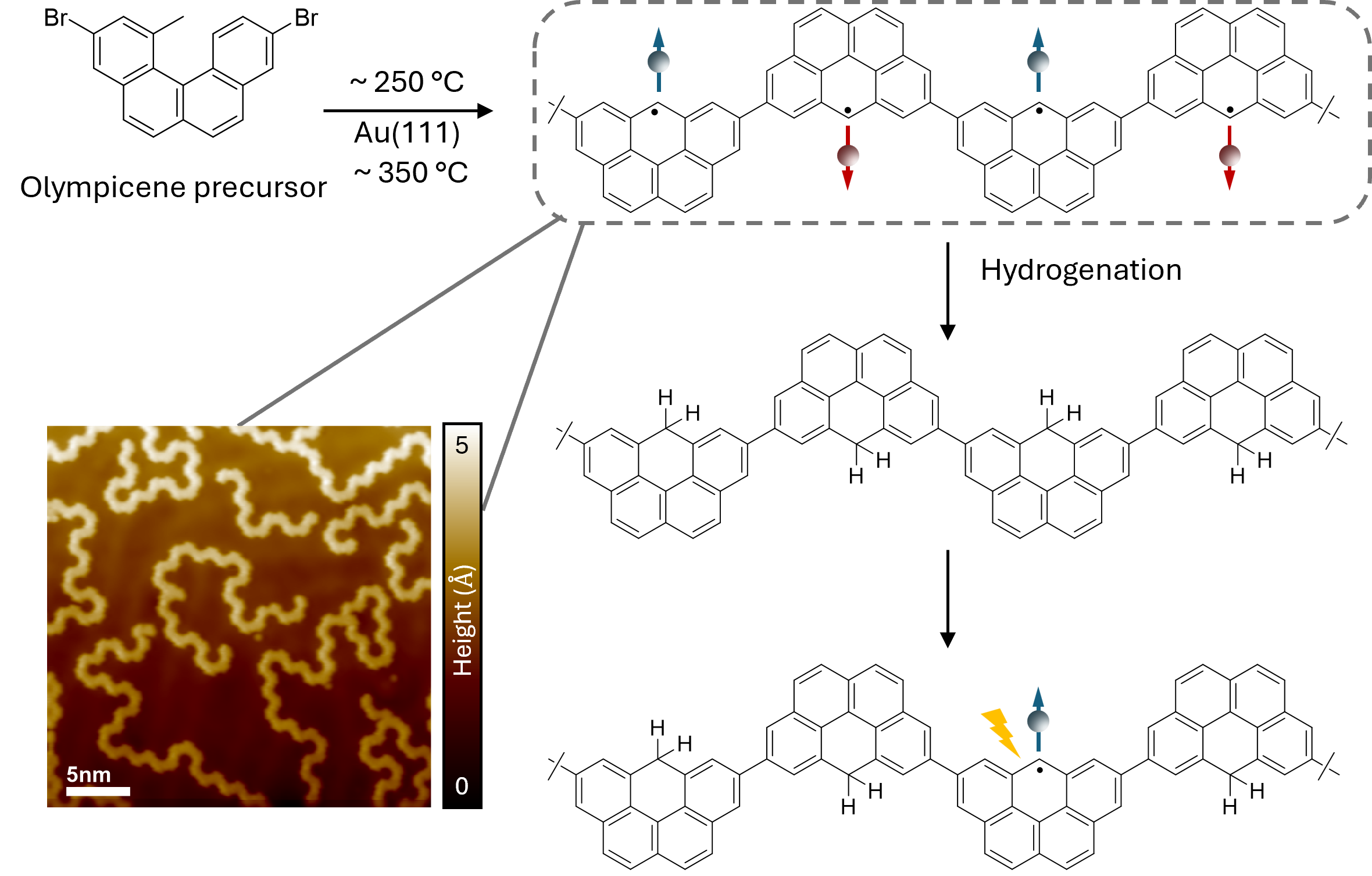}
\caption{On-surface synthesis route used to fabricate the Olympicene-based Heisenberg chains studied in this work. The STM image shows the Au(111) surface after the two-step annealing process. Imaging conditions: $V_{bias}=-100$~mV and $I_{set}=100$~pA.}
\label{fig:OSS}
\end{figure}

\subsection{On-surface synthesis}
\label{subsec:OSS}

Chain H was fabricated following the on-surface synthesis route shown in Fig.~\ref{fig:OSS}. Starting from the Olympicene precursor (chemical synthesis details in Sec.~\ref{sec:precursor}), the molecules were deposited onto a clean Au(111) surface and subsequently annealed in two stages to induce the formation of the nanographene spin chains. The resulting structures were then hydrogenated and selectively reactivated by tip-induced dehydrogenation, allowing controlled activation of individual spin sites.

\begin{figure}[h]
\centering
\includegraphics[width=\linewidth]{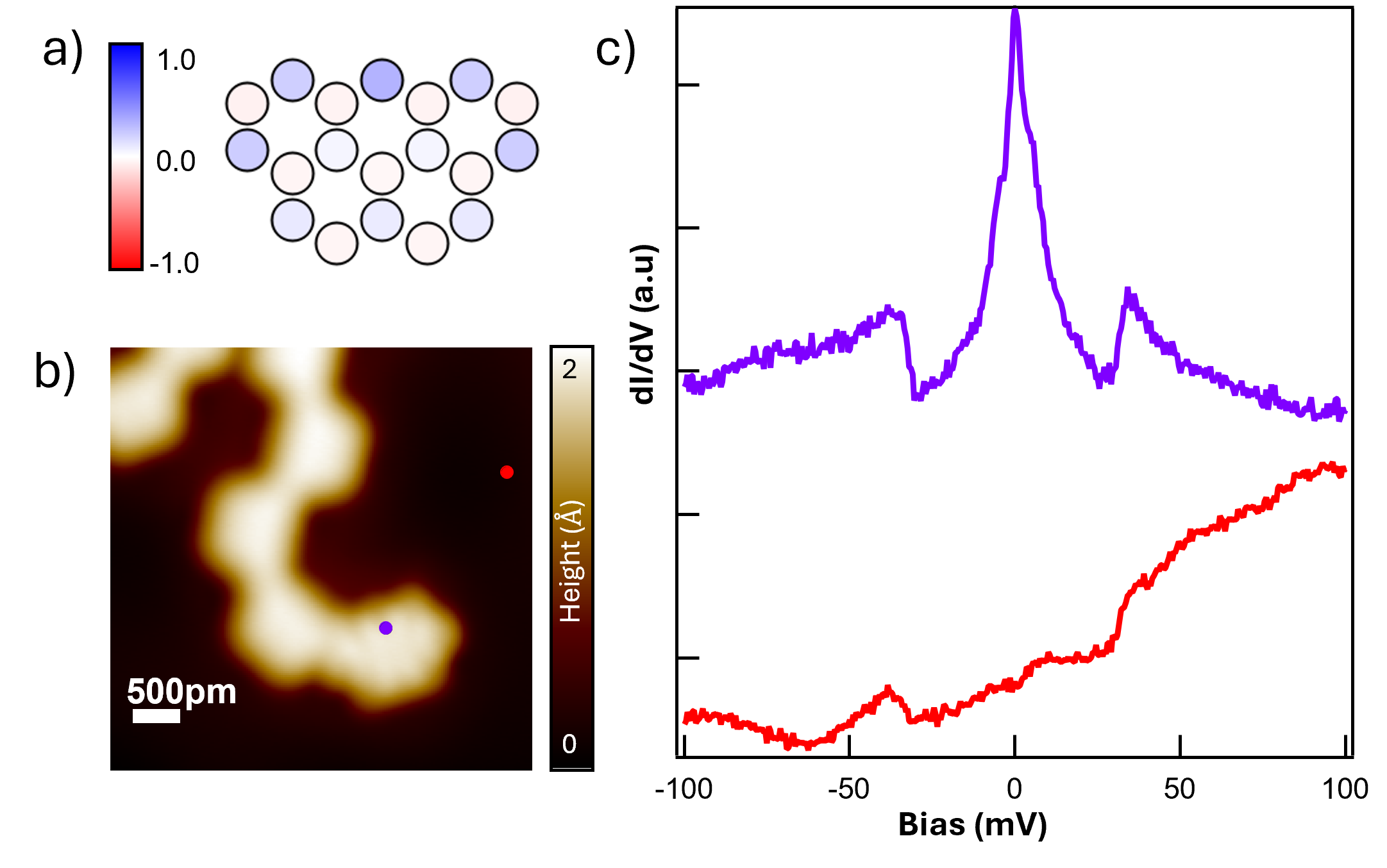}
\caption{(a) Spin density distribution of a single Olympicene unit obtained from mean-field Hubbard calculations. (b) STM image of chain with a singly activated Olympicene unit (purple marker). Imaging conditions: $V_{\mathrm{bias}}=-100~\mathrm{mV}$ and $I_{\mathrm{set}}=100~\mathrm{pA}$. (c) Representative $dI/dV$ spectra acquired at the positions indicated in (b). Spectroscopy conditions: $I_{\mathrm{set}}=1~\mathrm{nA}$ and $V_{\mathrm{mod}}=1~\mathrm{mV}$.
}
\label{fig:spectra}
\end{figure}

\subsection{$dI/dV$ spectroscopy}
\label{subsec:spectra}

STM and scanning tunneling spectroscopy (STS) measurements were performed in a low-temperature STM system (Scienta Omicron) operated at $T=4.5~\mathrm{K}$ and a base pressure below $1\times10^{-11}~\mathrm{mbar}$. All constant-current images and $dI/dV$ spectra were acquired using a CO-functionalized tip prepared by in-situ deposition of CO molecules.

Figure~\ref{fig:spectra} illustrates the spectroscopy protocol used throughout this work. The spin density distribution of an activated Olympicene molecule is shown in Fig.~\ref{fig:spectra}(a), together with an STM image of a chain with a singly activated unit in Fig.~\ref{fig:spectra}(b). The $dI/dV$ spectra [Fig.~\ref{fig:spectra}(c)] were acquired at the positions indicated by the colored markers, with spectroscopy on the Olympicene unit performed at the location of maximum spin density. All differential conductance measurements were obtained using lock-in detection with a modulation frequency of $691~\mathrm{Hz}$ and amplitude $V_{\mathrm{mod}}=1~\mathrm{mV}$.

\suppsection{Machine Learning Pipeline}
\label{sec:hamiltonian_learning}

In this section, we provide the technical details of the machine learning framework used to infer bond-resolved exchange interactions from STM spectroscopy. The goal of the model is to learn the mapping between local spectroscopic features and the underlying exchange couplings $J_n$ of an inhomogeneous Heisenberg Hamiltonian.
The approach is based on supervised learning using synthetic data generated from theoretical models, enabling controlled training and systematic benchmarking. Particular care is taken to ensure robustness with respect to experimental conditions, including limited bias windows, normalization procedures, and spectral broadening. The following subsections describe dataset generation, preprocessing, training methodology, and validation.

\subsection{Dataset Generation}
\label{subsec:dataset}

\begin{figure*}[htbp]
    \centering
    \includegraphics[width=\linewidth]{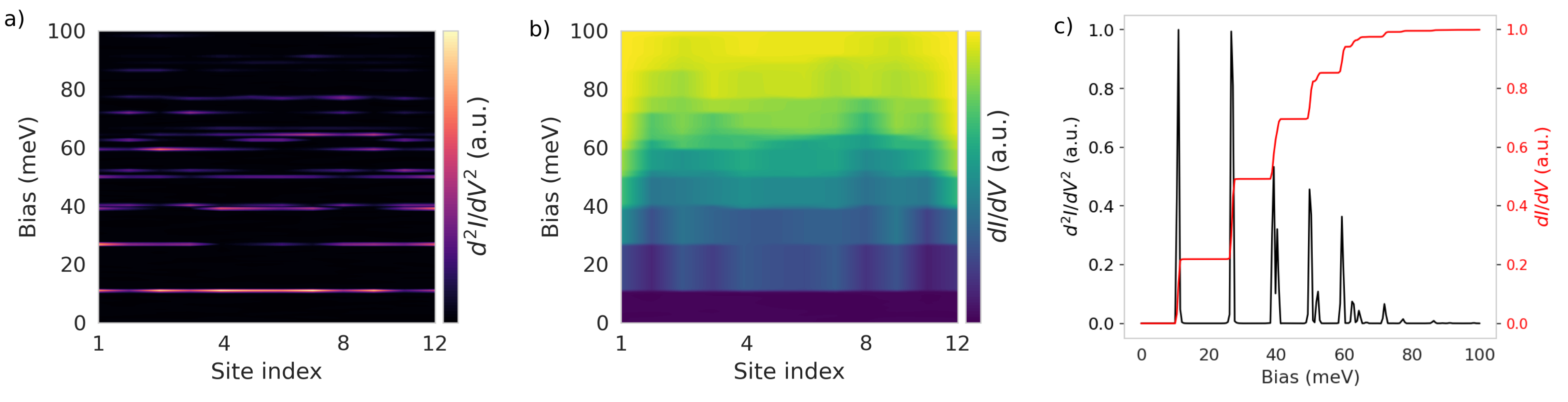}
    \caption{\textbf{Representative sample from the synthetic training dataset.} (a) $d^2I/dV^2$ and (b) the corresponding $dI/dV$ maps are shown as functions of site index and bias. Panel (c) shows an example of the $d^2I/dV^2$ and $dI/dV$ curves extracted at a representative site.}
    \label{fig:dataset}
\end{figure*}

The training dataset is generated from synthetic spectroscopic maps computed for inhomogeneous Heisenberg spin chains with randomly sampled exchange couplings $J_n$. Each realization consists of a chain of $12$ spin-$1/2$ sites described by the Hamiltonian
\begin{equation}
H = \sum_n J_n \, \mathbf{S}_n \cdot \mathbf{S}_{n+1}.
\end{equation}

The exchange couplings are randomly drawn from uniform distributions. In particular, we generate two sets of $1500$ chains each, with $J_n \in [35,45]$~meV and $J_n \in [30,40]$~meV, respectively. This combined dataset ensures coverage of the parameter range relevant for the experimental systems, while improving the robustness of the model against variations in the overall energy scale.

The first set is used to train a benchmark model based on ideal theoretical data, allowing us to assess the intrinsic reconstruction capability of the algorithm, as shown in Sec.~\ref{subsec:reconstruction}. The second set is then combined with the first to train a model, including experimental-like features, enabling a more realistic description of STM measurements [Sec.~\ref{subsec:experimental_features}].

For each sample, we compute the local dynamical spin response at site $n$ using a tensor network kernel polynomial algorithm~\cite{PhysRevB.83.195115,PhysRevResearch.1.033009,itensor,dmrgpy}. In general, this is defined as
\begin{equation}
A_n(\omega)=\sum_{\alpha=x,y,z}
\langle \Omega | {S_n^\alpha}
\delta(\omega-H+E_{\Omega})
S_n^\alpha | \Omega \rangle,
\end{equation}
where $|\Omega\rangle$ is the ground state with energy $E_{\Omega}$. In STM-based inelastic spectroscopy, the measured $d^2I/dV^2$ signal is proportional to this local dynamical response.

In the present work, we consider only the longitudinal contribution,
\begin{equation}
A_n^{zz}(\omega)=
\langle \Omega | {S_n^z}
\delta(\omega-H+E_{\Omega})
S_n^z | \Omega \rangle,
\end{equation}
which is sufficient since the Hamiltonian conserves total spin symmetry.

All training data are generated from even-length chains ($L=12$), whose ground state has total spin $S=0$. In this case, the longitudinal response $A_n^{zz}(\omega)$ fully captures the spin excitation spectrum.
For odd-length chains ($S=1/2$), computing the dynamical correlator requires accounting for the degeneracy of the ground state manifold. Even in this scenario,
it is worth noting that the \dIdV spectroscopy will not be given solely by the dynamical correlator, but in addition to it a zero bias Kondo 
peak will be present due to the ground state degeneracy. Such a zero bias anomaly is not captured by our methodology, which only accounts for second-order inelastic excitations. Our analysis will thus focus on even length chains when comparing with experiments, in which the Kondo channel is not relevant.
The resulting $d^2I/dV^2$ maps are computed for each site over a bias window of $[0, 100]$~meV sampled with 200 points.

While spin excitations are often analyzed experimentally in $d^2I/dV^2$, we use $dI/dV$ maps as input to the neural network [Fig.~\ref{fig:dataset}], as they provide a smoother representation of the spectra and are less sensitive to noise. The transformation is obtained by integrating along the bias axis,
\begin{equation}
\frac{dI}{dV}(V) \propto \int_0^{V} \frac{d^2 I}{dV'^2} dV'.
\end{equation}

The final dataset consists of spatially resolved spectroscopic maps, which are flattened and paired with the corresponding exchange couplings $\{J_n\}$, forming input-output pairs for supervised learning.

\subsection{Local inference via sliding windows}
\label{subsec:windows}

Rather than learning the full set of exchange couplings $\{J_n\}$ from an entire spectroscopic map, we adopt a local inference strategy based on sliding windows, following Ref.~\cite{Koch2025}. This approach exploits the fact that the local spectroscopic response is primarily sensitive to exchange interactions in a limited spatial region.
Starting from spatially resolved $dI/dV$ maps of chains with $L=12$ sites and $200$ bias points per site, we construct input features by selecting contiguous windows of three neighboring sites. Each window is flattened into a feature vector of dimension $3 \times 200 = 600$, corresponding to the concatenated spectra of the three sites.
For each window centered on sites $(n,n+1,n+2)$, the target labels are the two exchange couplings $(J_n, J_{n+1})$. By sliding the window across the chain, a single spectroscopic map yields multiple input-output pairs, increasing the size of the training dataset.
This local formulation allows the neural network to learn short-range correlations between spectral features and exchange interactions. Moreover, since the prediction is based on local information, the trained model can be applied to chains of arbitrary length.
To avoid information leakage between training and test sets, the data splitting is performed at the level of full spectroscopic maps, and the sliding windows are constructed separately within each subset.

\subsection{Reconstruction of exchange couplings for arbitrary chain lengths}
\label{subsec:reconstruction}

\begin{figure}[htbp]
    \centering
    \includegraphics[width=\linewidth]{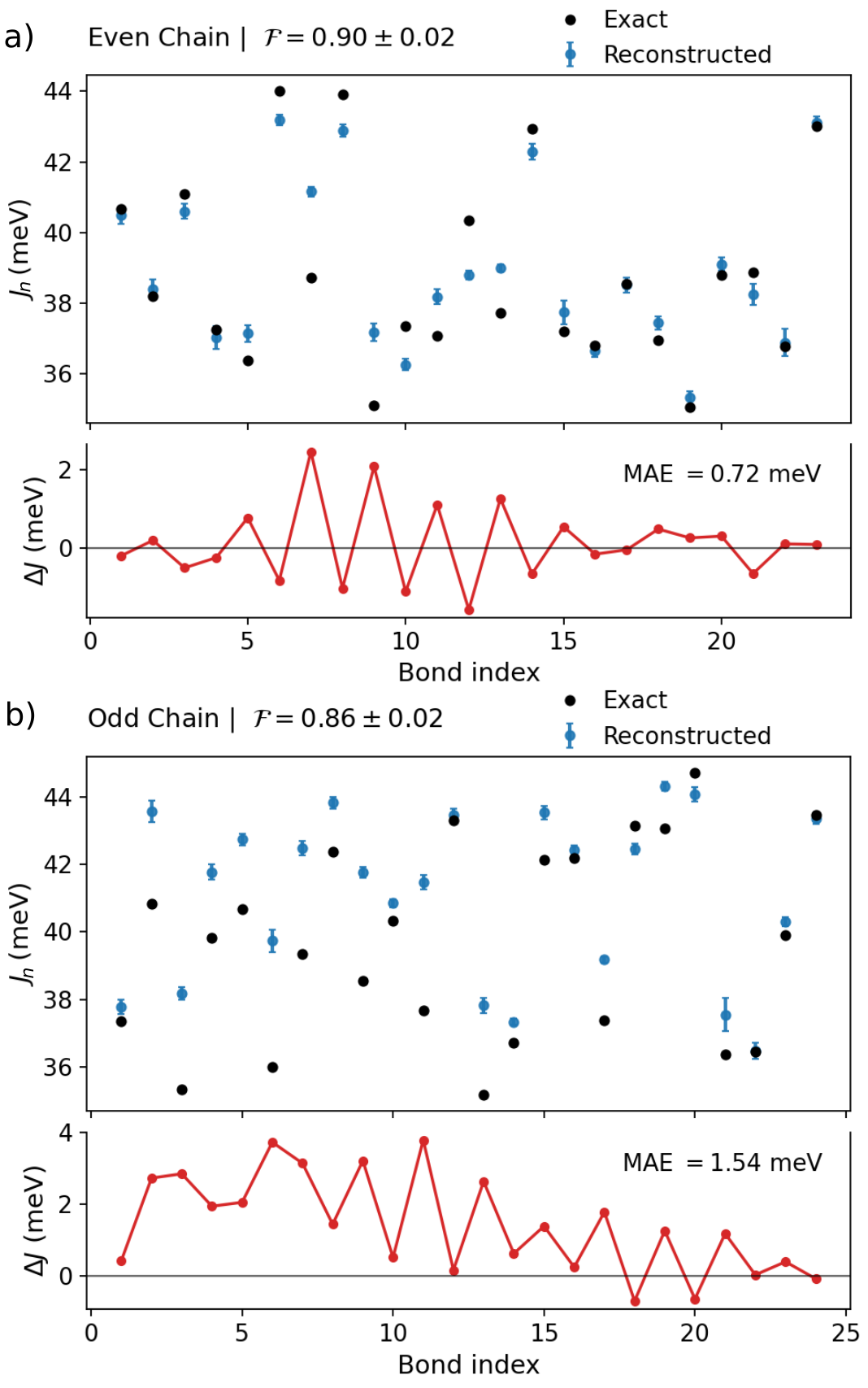}
    \caption{Reconstruction of the exchange couplings for (a) an even ($L=24$) and (b) an odd ($L=25$) chain. The upper panels show the exact and reconstructed exchange profiles, while the lower panels display the bond-resolved error $\Delta J = J_{\mathrm{pred}} - J_{\mathrm{true}}$. The reported MAE is computed over all bonds in the chain.}
    \label{fig:reconstruction}
\end{figure}

To assess the intrinsic capability of the reconstruction scheme, we first consider a model trained exclusively on ideal theoretical data, without including experimental distortions. In this setting, the exchange couplings are sampled in the range $J \in [35,45]$~meV, and the input spectra correspond to integrated theoretical simulations.
Once trained, the algorithm is applied to full chains by sliding the same three-site window along the system. For an open chain of length $L$, this produces $L-2$ overlapping predictions, each providing estimates for two consecutive exchange couplings.
Specifically, for each window centered on sites $(n,n+1,n+2)$, the network outputs a pair $(J_n, J_{n+1})$. The full set of couplings $\{J_n\}$ is reconstructed by combining these overlapping predictions. The edge couplings are taken directly from the first and last windows, while bulk couplings are obtained by averaging the two estimates coming from adjacent windows,
\begin{equation}
J_n = \frac{J_n^{(n-1)} + J_n^{(n)}}{2} ,
\end{equation}
where $J_n^{(n-1)}$ and $J_n^{(n)}$ denote the predictions from the windows $(n-1,n,n+1)$ and $(n,n+1,n+2)$, respectively.

This overlap-averaging procedure reduces local prediction noise and yields a consistent reconstruction of the full exchange profile. To further improve robustness, we employ an ensemble of $10$ independently trained neural networks with different random initializations. The final prediction is obtained by averaging over the ensemble, and the associated uncertainty is estimated from the standard deviation across models.

We validate this approach by applying the model to chains longer than those used during training [Fig.~\ref{fig:reconstruction}]. To quantify the accuracy of the reconstruction, we use the fidelity $\mathcal{F}$, defined as the normalized correlation between predicted and true exchange couplings (see Sec.~\ref{sec:fidelity} for details). A value $\mathcal{F}=1$ corresponds to perfect agreement.

To complement the fidelity metric, we also report the mean absolute error (MAE) of the reconstructed exchange couplings,
\begin{equation}
\mathrm{MAE_{J_n}}
=
\frac{1}{N_b}
\sum_{b=0}^{N_b}
\left|
J_b^{\mathrm{pred}}
-
J_b^{\mathrm{exact}}
\right|,
\end{equation}
where $N_b = L-1$ is the number of bonds in the chain. The MAE provides a direct measure of the average bond-resolved reconstruction error in units of meV.

For an even chain with $L=24$, the reconstructed exchange profile closely follows the exact couplings, demonstrating generalization beyond the training size with $90 \%$ accuracy. For an odd chain with $L=25$, the fidelity is reduced to $86 \%$, consistent with the different low-energy spectral structure of odd spin-$1/2$ chains. In particular, odd chains exhibit spectral weight at zero bias associated with the ground-state doublet, whereas the training set consists exclusively of gapped even-length chains with total spin $S=0$. Although the integration to $dI/dV$ reduces the prominence of this zero-bias feature, the underlying low-energy structure remains qualitatively different from that encountered during training. The corresponding mean absolute reconstruction errors are $0.72$ meV for the even chain and $1.54$ meV for the odd chain, confirming that the predicted exchange couplings remain quantitatively close to the exact values in both cases.

\subsection{Experimental-like preprocessing}
\label{subsec:experimental_features}

To bridge the gap between theoretical simulations and experimental STM data, we progressively introduce realistic features into the training dataset. In particular, we (i) extend the distribution of exchange couplings to a wider range ($J_n \in [30,40]$~meV and $J_n \in [35,45]$~meV), to increase the diversity of the training set while preserving physically reasonable variations between neighboring bonds, (ii) include noise, broadening, and background contributions, and (iii) restrict the bias window to experimentally accessible energies.

\begin{table}[htbp]
\centering
\caption{Parameters used to generate experimentally enhanced spectra.}
\begin{tabular}{lc}
\hline
Parameter & Value \\
\hline
Offset & $\mathcal{U}(0.009,0.016)$ \\
Lorentzian width $\gamma$ & $\mathcal{U}(0.5,2.0)$ \\
Bias cutoff & $50$ meV \\
Resampled bias points & $200$ \\
Noise amplitude $\sigma_0$ & $0.002$ \\
Noise weight $w(V)$ & linear from $0.6$ to $1.2$ \\
Baseline subtraction window & $0$--$3$ meV \\
Normalization window & $40$--$50$ meV \\
Drift slope & $\mathcal U(4.8\times10^{-5},1.5\times10^{-4})$ \\
\hline
\end{tabular}
\label{tab:augmentation}
\end{table}

\subsubsection{Spectral transformations}
To mimic experimental STM spectra, each simulated \dIdV at site $n$ is transformed through a sequence of randomized augmentations.
First, a constant offset is added, sampled uniformly in the range $[0.009,0.016]$. The spectra are then broadened by convolution with a Lorentzian kernel
\begin{equation}
K(x)=\frac{1}{1+(x/\gamma)^2},
\end{equation}
normalized to unit weight, where the broadening parameter is sampled uniformly in the range $\gamma\in[0.5,2.0]$. The bias range is subsequently truncated to $0$--$50$ meV and resampled to 200 points.

To emulate instrumental noise, Gaussian fluctuations are added at each bias point $i$ according to
\begin{equation}
\left(\frac{dI}{dV}\right)_i^{(n)}  \rightarrow \left(\frac{dI}{dV}\right)_i^{(n)}  + \epsilon_i,
\qquad
\epsilon_i \sim \mathcal N\left(0,\sigma_i^2\right),
\end{equation}
with a bias-dependent standard deviation
\begin{equation}
\sigma_i = \sigma_0 w_i ,
\end{equation}
where $\sigma_0=0.002$ and $w_i$ increases linearly from $0.6$ to $1.2$ across the bias window.

\begin{figure}[htbp]
    \centering
    \includegraphics[width=0.8\linewidth]{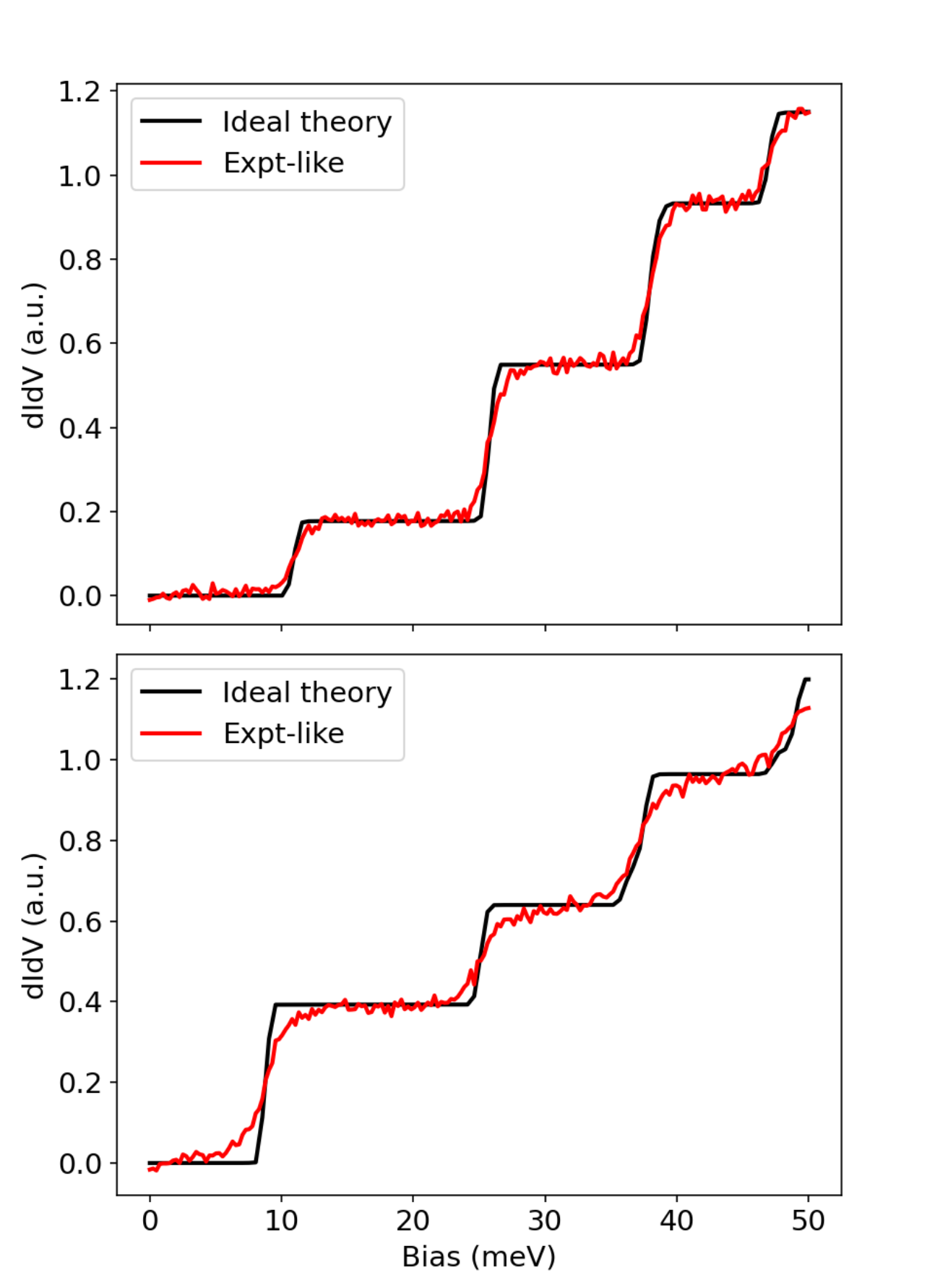}
    \caption{Two representative $dI/dV$ curves before and after experimental-like preprocessing, illustrating the effect of varying Gaussian noise levels and Lorentzian broadening strengths.}
    \label{fig:expt}
\end{figure}

A weak linear drift is further introduced at higher bias,
\begin{equation}
\left(\frac{dI}{dV}\right)_i^{(n)}  \rightarrow \left(\frac{dI}{dV}\right)_i^{(n)}  + s(i-i_0),
\end{equation}
for $i>i_0$, where the onset point $i_0$is chosen according to the spectral amplitude, ensuring that the drift primarily affects the higher-bias region, and the slope $s$ is randomly sampled in the range $[4.8\times10^{-5},1.5\times10^{-4}]$.
Finally, each \dIdV curve is normalized by subtracting the average signal below $3$ meV and dividing by the mean absolute intensity in the interval $40$--$50$ meV. This ensures that the neural network focuses on relative spectral features rather than their absolute magnitude.

These transformations, summarized in Table~\ref{tab:augmentation}, generate synthetic data that closely resemble experimental STM spectroscopy, improving the robustness and transferability of the trained models. Two representative $dI/dV$ curves before and after the spectral transformations are shown in Fig.~\ref{fig:expt}.

\subsection{Bias cutoff and spectral mismatch}
\label{subsec:cutoff}

\begin{figure}[htbp]
    \centering
    \includegraphics[width=\linewidth]{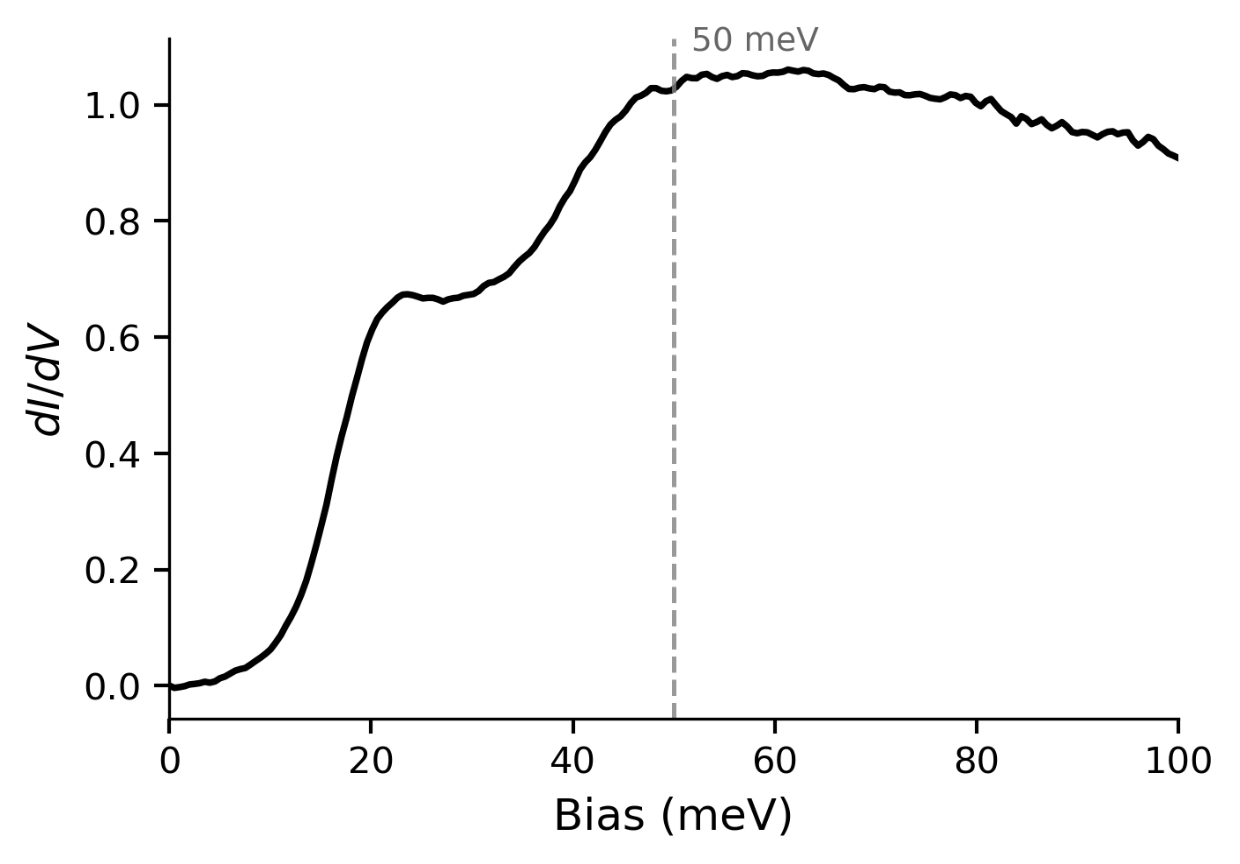}
    \caption{Experimental $dI/dV$ curve illustrating the characteristic decay of the signal beyond a $\sim 50$~meV bias.}
    \label{fig:cutoffs}
\end{figure}

The choice of the bias window used for training plays a crucial role in the performance of the reconstruction. While theoretical spectra extend to arbitrarily high energies, experimental $dI/dV$ signals exhibit a progressive loss of spectral weight at large bias together with background contributions that are not captured by the ideal Heisenberg model, as shown in Fig.~\ref{fig:cutoffs}.
At high bias, the experimental signal is increasingly dominated by non-universal effects, including background decay and tunneling contributions beyond the scope of the model. Incorporating this energy range into the training data therefore introduces systematic discrepancies between theory and experiment, forcing the neural network to fit features that are unrelated to the underlying exchange interactions. As a result, the reconstruction accuracy deteriorates, particularly for the low-energy part of the spectrum that carries the relevant information on the exchange couplings.

Based on this analysis, we select a cutoff of $50$~meV as the optimal compromise between retaining sufficient spectral information and minimizing the impact of high-energy experimental artifacts. This choice is further supported by the high reconstruction fidelity obtained on synthetic experimental-like test data (see Sec.~\ref{subsec:noise}) and by the good agreement with experimental measurements, as demonstrated by the reconstructed spectroscopic maps and the MAE reported in Sec.~\ref{sec:error}.

\subsection{Fidelity Evaluation}
\label{sec:fidelity}

To assess the performance of the trained algorithm, we define the fidelity $\mathcal{F}$ as the normalized correlation between predicted and true exchange couplings. Since each three-site window yields two couplings, $(J_n, J_{n+1})$, the fidelity is computed over both components:

\begin{equation}
\mathcal{F} = 
\frac{
\left\langle \mathbf{J}^{\mathrm{pred}} \cdot \mathbf{J}^{\mathrm{true}} \right\rangle 
- 
\left\langle \mathbf{J}^{\mathrm{pred}} \right\rangle 
\cdot 
\left\langle \mathbf{J}^{\mathrm{true}} \right\rangle
}{
\sqrt{\mathrm{var}(\mathbf{J}^{\mathrm{pred}})} 
\sqrt{\mathrm{var}(\mathbf{J}^{\mathrm{true}})}
},
\end{equation}
where $\mathbf{J} = (J_n, J_{n+1})$ denotes the pair of exchange couplings associated with a given window, and $\langle \cdot \rangle$ indicates averaging over the dataset.

The fidelity takes values in the interval $[0,1]$, where $\mathcal{F}=1$ corresponds to perfect agreement between predictions and ground truth, while $\mathcal{F}=0$ indicates no correlation.

\subsection{Neural network architecture}
\label{sec:nn_arch}

The local inverse mapping from a three-site spectroscopic window to the two corresponding exchange couplings is learned using a feedforward neural network (FFNN). Each input consists of a flattened $3 \times 200$ local $dI/dV$ window, resulting in an input dimension of $600$. The output layer has two neurons, corresponding to the predicted exchange couplings $(J_n,J_{n+1})$.

The local exchange couplings are normalized according to
\begin{equation}
\tilde{J}_n = \frac{J_n - J_{\mathrm{min}}}{J_{\mathrm{max}} - J_{\mathrm{min}}},
\end{equation}
where $J_{\mathrm{min}} = 30$~meV and $J_{\mathrm{max}} =45$~meV correspond to the minimum and maximum exchange couplings within the whole training dataset.

\begin{figure}[htbp]
    \centering
    \includegraphics[width=\linewidth]{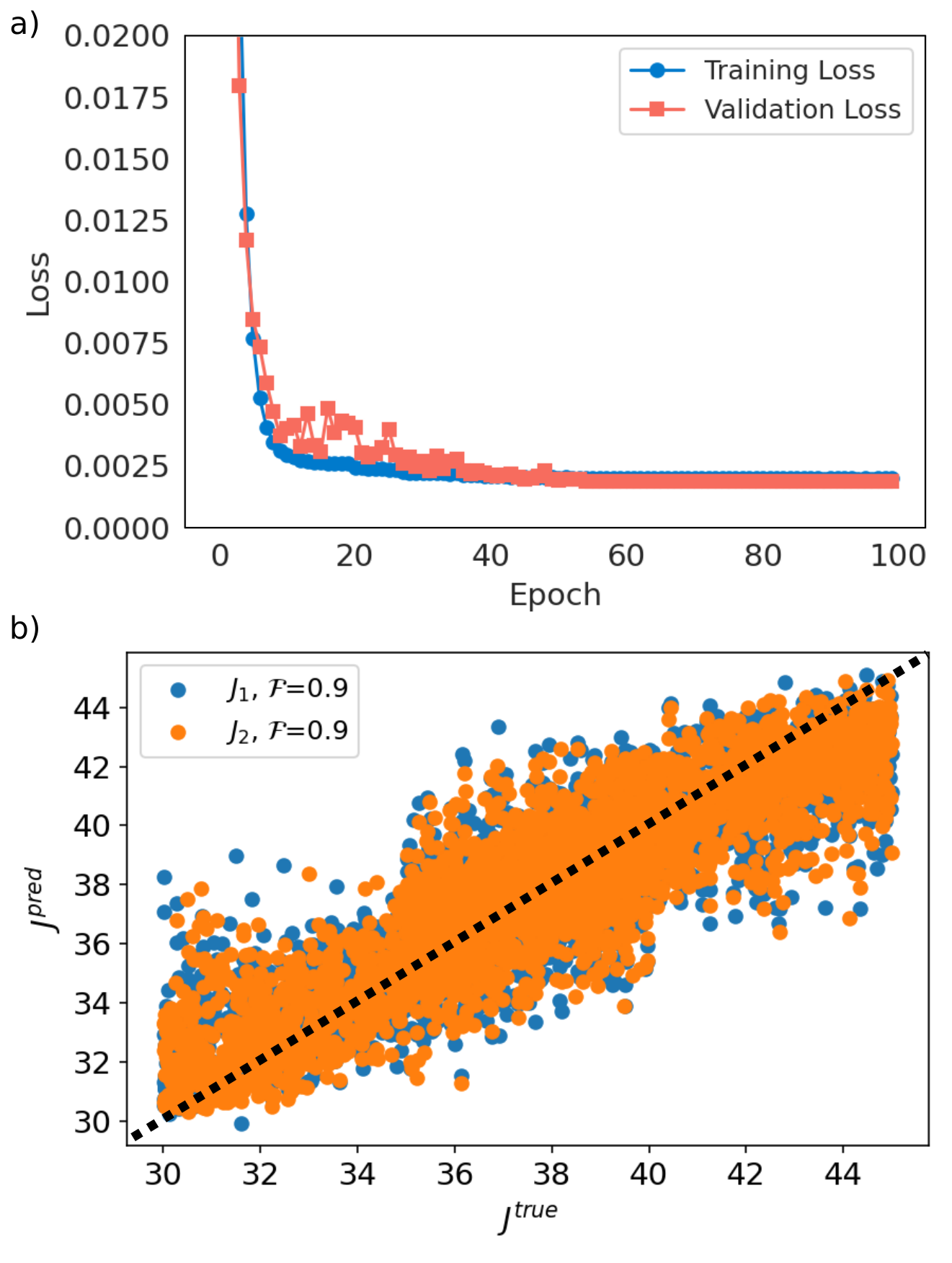}
    \caption{(a) Training and validation loss curves and (b) predictions of a representative model from the ensemble trained with a $50$~meV cutoff and experimental-like preprocessing. The predicted exchange couplings $J_1$ and $J_2$ correspond to the bonds between sites $(1,2)$ and $(2,3)$ within each three-site sub-window. The dotted line indicates perfect predictions.}
    \label{fig:loss_curves}
\end{figure}

The network is trained using the Huber loss, which provides robustness against outliers in the target exchange couplings. L2 regularization is applied to all dense layers to reduce overfitting. Training is performed with early stopping based on the validation loss, with a patience of 12 epochs and restoration of the best weights. In addition, the learning rate is reduced by a factor of 0.5 if the validation loss does not improve for 4 epochs, down to a minimum learning rate of $10^{-6}$.
The training of the model with a $50$~meV cutoff does not exhibit signs of overfitting, as evidenced by the convergence of training and validation losses shown in Fig.~\ref{fig:loss_curves}a. To further assess the predictive performance, we evaluate the model on an independent test set. The comparison between predicted and true exchange couplings demonstrates agreement up to $90$\% fidelity, as shown in Fig.~\ref{fig:loss_curves}b.
The network architecture and training hyperparameters are summarized in Table~\ref{tab:nn_J}.
The notation Dense($N$) denotes a fully connected layer with $N$ units, BatchNorm denotes batch normalization, and Dropout($p$) indicates dropout regularization with probability $p$. ReLU activations are used in the hidden layers, while the output layer is linear.
\begin{table}[htbp]
\centering
\caption{Neural network architecture used for local exchange coupling reconstruction.}
\label{tab:nn_J}
\begin{tabular}{@{}ll@{}}
\toprule
\textbf{Layer (Type)} & \textbf{Output dimension} \\
\midrule
Input & (600) \\
Dense(512) + ReLU & (512) \\
BatchNorm & (512) \\
Dropout(0.25) & (512) \\
Dense(256) + ReLU & (256) \\
BatchNorm & (256) \\
Dropout(0.25) & (256) \\
Dense(128) + ReLU & (128) \\
Dense(2) & (2) \\
\midrule
\textbf{Optimizer} & Adam, learning rate $3\times10^{-4}$ \\
\textbf{Loss function} & Huber, $\delta=0.02$ \\
\textbf{Metric} & Mean absolute error \\
\textbf{L2 regularization} & $\lambda = 3\times10^{-4}$ \\
\textbf{Batch size} & 128 \\
\textbf{Maximum epochs} & 100 \\
\bottomrule
\end{tabular}
\end{table}

We also explored one- and two-dimensional convolutional neural networks (CNNs) as alternative architectures. While CNNs achieved predictive fidelities comparable to those of the FFNN, they did not provide a systematic improvement across different training runs. This behavior is expected, as the input consists of a small three-site spectroscopic window, for which the relevant information is already contained in a low-dimensional representation. In contrast to image-recognition tasks, the present problem does not exhibit large-scale translational features that would strongly benefit from convolutional filters. We therefore adopted the simpler FFNN architecture throughout this work.

\subsection{Noise robustness analysis}
\label{subsec:noise}

To assess the robustness of the learning framework against experimental noise, we evaluate the performance of the trained models under controlled perturbations of the input spectra. In particular, we compare two models introduced above: a benchmark model trained on ideal theoretical data and a model trained on datasets incorporating experimental-like preprocessing, including noise, broadening, background contributions and a bias cutoff of $50$~meV.

Noise is introduced at the level of the $dI/dV$ maps by adding a random perturbation to each data point,
\begin{equation}
\left(\frac{dI}{dV}\right)^{\mathrm{noisy}}_{ij} =\left(\frac{dI}{dV}\right)_{ij}
+
\varepsilon_{ij},
\qquad
\varepsilon_{ij} \sim \mathcal{U}(-\eta,\eta),
\end{equation}
where $\mathcal{U}(-\eta,\eta)$ denotes a uniform distribution of width $2\eta$. The noise amplitude is defined relative to the typical signal intensity as
\begin{equation}
\eta = \alpha A,
\end{equation}
where $A$ is the characteristic amplitude of the corresponding $dI/dV$ spectrum and $\alpha$ represents the relative noise level. This choice ensures zero-mean noise while allowing a consistent comparison between datasets with different normalization schemes.
For each noise level, the fidelity $\mathcal{F}$ (defined in Sec.~\ref{sec:fidelity}) is computed on the corresponding test set. To reduce statistical fluctuations, we employ an ensemble of $10$ independently trained neural networks and report the mean fidelity together with the corresponding standard deviation.

\begin{figure}[htbp]
    \centering
    \includegraphics[width=\linewidth]{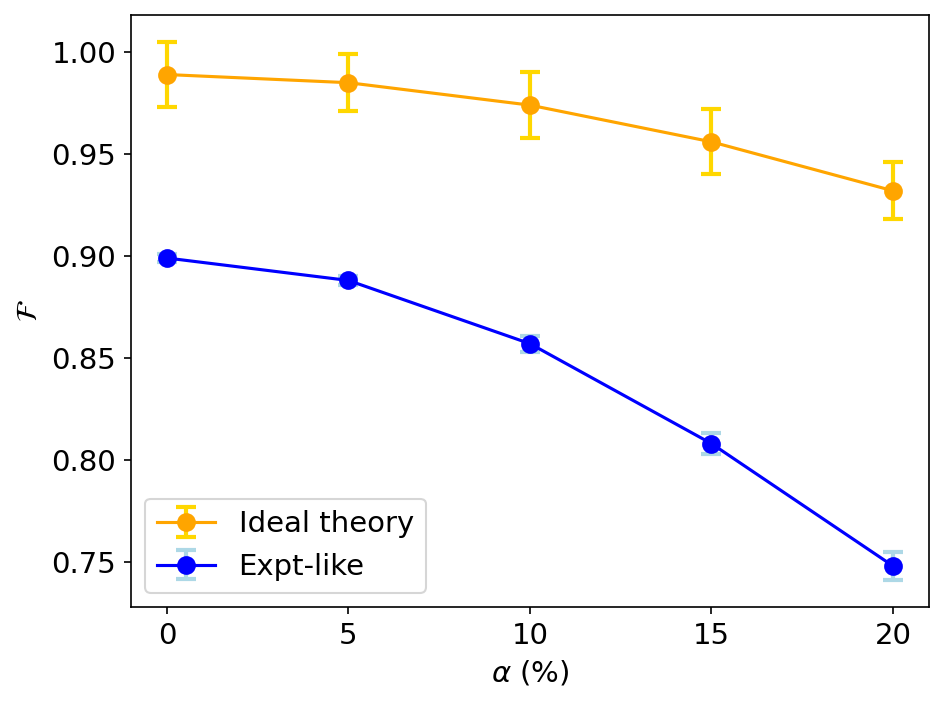}
    \caption{Fidelity as a function of the relative noise strength $\alpha$ for ensembles of models trained (and tested) on ideal theoretical data and on datasets including experimental-like preprocessing.}
    \label{fig:noise_test}
\end{figure}

As shown in Fig.~\ref{fig:noise_test}, the benchmark model trained on ideal theoretical data achieves higher fidelity over the full range of noise levels considered. It should be emphasized, however, that the two curves are evaluated on different test sets and are therefore not intended as a direct comparison between training strategies. Especially, the model trained with experimental-like preprocessing is tested on spectra that already include broadening, normalization, background contributions, and restricted bias windows. Nevertheless, the fidelity remains high and comparatively stable under increasing noise, demonstrating that the reconstruction retains strong predictive performance even in the presence of experimental-like distortions. While such preprocessing reduces the ideal case reconstruction accuracy, it is essential for improving the transferability of the framework to experimentally relevant STM spectra.

\suppsection{Experimental inference and validation}

In this section, we apply the Hamiltonian learning framework to experimental STM spectroscopy of olympicene spin chains. We first demonstrate the reconstruction of spatially resolved spectroscopic maps for the representative systems, highlighting both inhomogeneous and near-uniform regimes.
To test the generality of the approach, we analyze additional chains beyond the main examples discussed in the main text. These chains were synthesized and measured using the same experimental protocol as Chain IH in Ref.~\cite{Zhao2025}, allowing us to assess whether the inferred exchange inhomogeneity is a generic feature of this class of nanographene spin chains.
Finally, we provide a quantitative validation of the reconstruction by comparing experimental and predicted spectroscopic maps, including an analysis of the MAE. Together, these results demonstrate the accuracy, robustness, and broad applicability of the method to experimental data.

\subsection{Reconstruction of experimental spectroscopic maps}
\label{sec:maps}

\begin{figure*}[t]
    \centering
    \includegraphics[width=\linewidth]{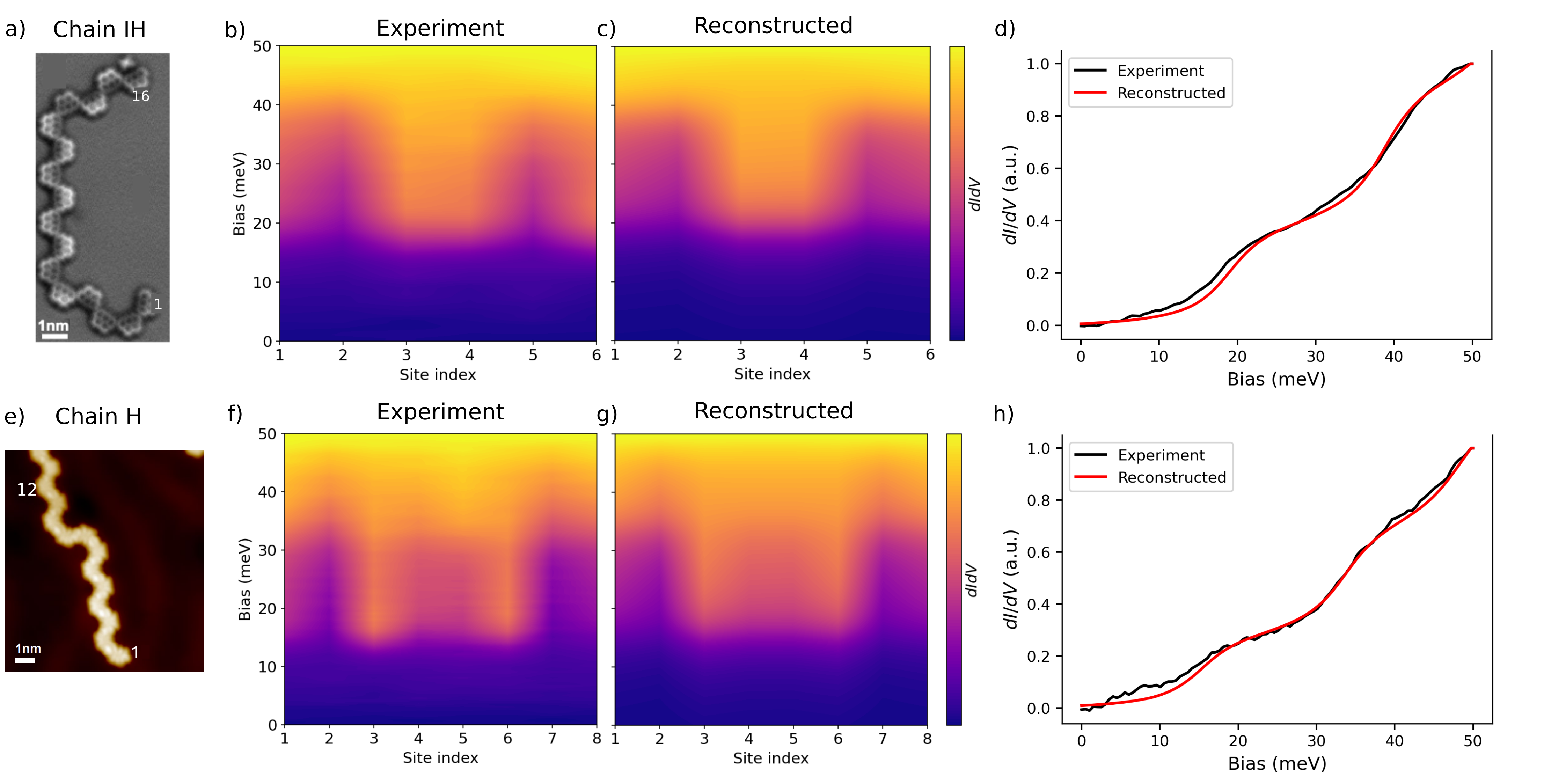}
    \caption{
    \textbf{Reconstruction of experimental spectroscopic maps.}
    (a,e) Structural microscopy images of Chain IH and Chain H;
    Representative experimental $dI/dV$ maps for (b) Chain IH ($L=$~6) and (f) Chain H ($L=$~8);
    (c,g) Reconstructed maps from the inferred inhomogeneous Heisenberg Hamiltonian; 
    (d,h) Comparison of representative local $dI/dV$ spectra, showing good agreement between experiment and reconstruction.
    }
    \label{fig:chainAB}
\end{figure*}

We validate the performance of the Hamiltonian learning framework by reconstructing spatially resolved spectroscopic maps from experimental STM data. Figure~\ref{fig:chainAB} shows representative results for two systems introduced in the main: Chain IH, which exhibits pronounced inhomogeneity, and Chain H, which lies close to the homogeneous limit. Structural microscopy images are shown in Fig.~\ref{fig:chainAB}a and Fig.~\ref{fig:chainAB}e.
The experimental spectroscopy is initially acquired over a bias window ranging from $-100$~meV to $100$~meV. To match the symmetry of the theoretical framework and reduce experimental asymmetries, we perform the inference on symmetrized maps obtained as the average between positive and negative bias signals. The resulting maps are then processed using the same normalization, scaling and bias cutoff procedures employed during training.
For each system, we compare the experimental $dI/dV$ maps with those obtained from the reconstructed inhomogeneous Heisenberg Hamiltonian [Fig.~\ref{fig:chainAB}b-c and Fig.~\ref{fig:chainAB}f-g]. In both cases, the model accurately reproduces the main spectral features, including the spatial dependence of the excitation energies. The agreement is particularly evident in the evolution of the low energy steps.
To better visualize the reconstruction quality, we also compare representative local $dI/dV$ curves. As shown in Fig.~\ref{fig:chainAB}d and Fig.~\ref{fig:chainAB}h, the reconstructed curves closely follow the experimental ones, capturing both the position and relative height of the spectroscopic features.
These results demonstrate that the inferred bond-resolved exchange couplings provide a consistent and physically meaningful description of the experimental data, enabling a faithful reconstruction of the observed spectroscopic response in both inhomogeneous and near-uniform regimes.



%
%

\begin{figure*}[t]
    \centering
    \includegraphics[width=\linewidth]{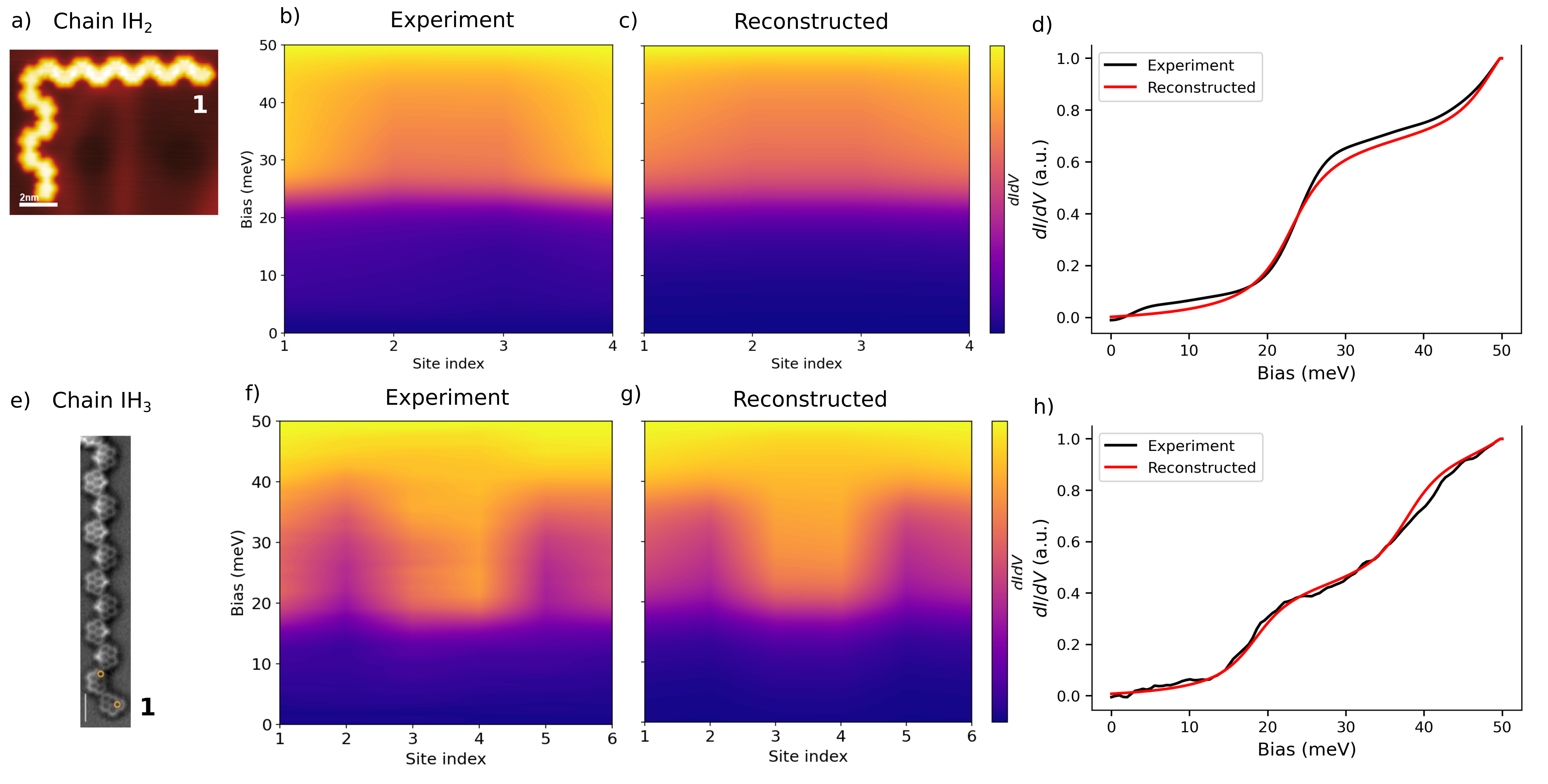}
    \caption{
    \textbf{Reconstruction of experimental spectroscopic maps.}
    (a,e) Structural microscopy images of Chain IH$_2$ and Chain IH$_3$;
    Representative experimental $dI/dV$ maps for (b) Chain IH$_2$ ($L=$~4) and (f) Chain IH$_3$ ($L=$~6);
    (c,g) Reconstructed maps from the inferred inhomogeneous Heisenberg Hamiltonian; 
    (d,h) Comparison of representative local $dI/dV$ spectra, showing good agreement between experiment and reconstruction.
    }
    \label{fig:chain_CD}
\end{figure*}

To further assess the generality of the reconstruction, we apply the same procedure to additional olympicene spin chains synthesized and characterized following the experimental protocol of Ref.~\cite{Zhao2025}. Here we denote the Chain IH as Chain IH$_1$ and label the additional chains as Chain IH$_2$ and Chain IH$_3$. These systems provide an independent test of whether exchange inhomogeneity is a recurring feature of this class of nanographene spin chains.
The reconstructed spectroscopic maps for these chains show a level of agreement with the experimental data similar to that observed for Chain IH$_1$ [Fig.~\ref{fig:chain_CD}]. In particular, the model consistently captures the spatial evolution of the excitation features, confirming that the inferred exchange profiles provide a robust description of this class of systems.

\begin{figure*}
    \centering
    \includegraphics[width=1\linewidth]{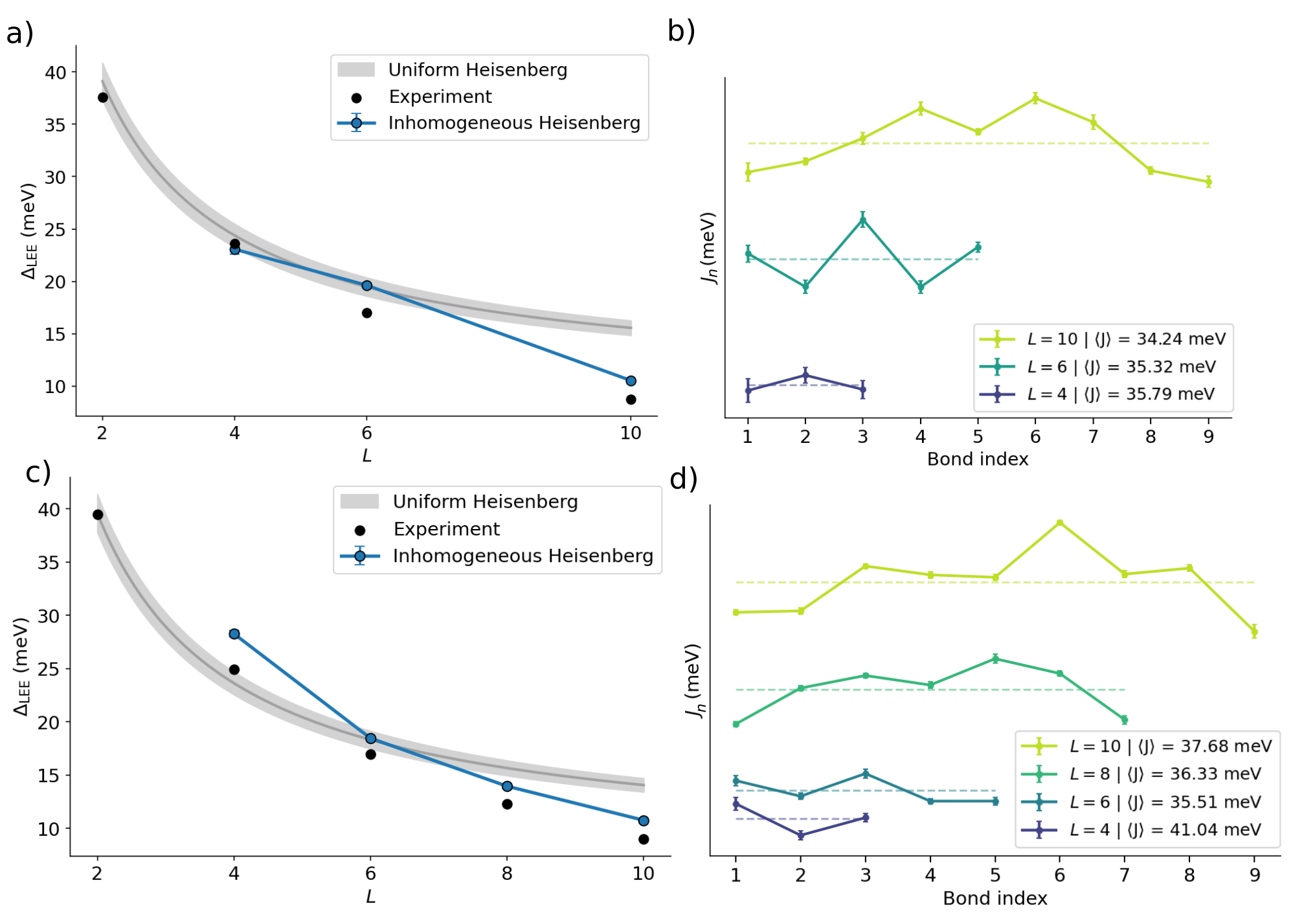}
    \caption{
    \textbf{Reconstruction results for additional olympicene spin chains.} 
    (a,c) Experimental low-energy excitation gap $\Delta_{\mathrm{LEE}}$ as a function of chain length $L$, compared with the predictions of uniform and inhomogeneous Heisenberg models, for (b) Chain IH$_2$ and (c) Chain IH$_3$. 
    (b,d) Bond-resolved exchange couplings $J_n$ reconstructed for Chains IH$_2$ and IH$_3$ at different chain lengths. Dashed lines indicate the average exchange coupling $\langle J \rangle$ for each chain length. The reconstructed profiles reveal significant spatial variations of the exchange interaction, consistent with the behavior observed in Chain IH$_1$.
    }
    \label{fig:extra_chains}
\end{figure*}

We further compare the evolution of the low-energy excitation gap as a function of chain length with the exchange couplings reconstructed for the different chains [Fig.~\ref{fig:extra_chains}]. Similarly to Chain IH$_1$, the additional chains exhibit clear bond-dependent variations of the exchange interaction.
Quantitatively, the inhomogeneous reconstruction systematically improves the agreement with the experimental excitation gaps compared to a uniform Heisenberg description. To quantify this agreement, we define the MAE between the experimental and predicted gaps as

\begin{equation}
\label{eq:mae_gap}
\mathrm{MAE_{\Delta}}
=
\frac{1}{N_L}
\sum_{i=1}^{N_L}
\left|
\Delta_i^{\mathrm{model}}
-
\Delta_i^{\mathrm{exp}}
\right|,
\end{equation}
where $\Delta_i^{\mathrm{exp}}$ and $\Delta_i^{\mathrm{model}}$ denote the experimental and theoretical excitation gaps for a chain of length $L_i$, respectively, and $N_L$ is the number of chain lengths considered. The error is evaluated separately for the inhomogeneous and uniform Heisenberg descriptions by substituting the corresponding theoretical predictions into Eq.~(\ref{eq:mae_gap}). For Chains IH$_2$ and IH$_3$, the error is reduced from $2.91$ meV to $1.24$ meV and from $2.00$ meV to $1.65$ meV, respectively, further demonstrating the robustness and transferability of the framework across distinct experimental realizations.

\begin{table}[h!]
\centering
\caption{MAE between experimental and reconstructed spectroscopic maps for different effective chain lengths $L$. Values are reported in \%. The last column shows the average MAE over all available lengths for each chain.}
\label{tab:mae_maps}
\begin{tabular}{@{}lccccccr@{}}
\toprule
\textbf{Chain} 
& \multicolumn{6}{c}{$L$} 
& \textbf{Avg.} \\
\cmidrule(lr){2-7}
& 4 & 6 & 8 & 10 & 12 & 16 & \\
\midrule

H
& $ 6.1$
& $ 4.3$
& $2.9$
& $ 3.7$
& $ 4.4$
& --
& $\mathbf{4.3}$ \\

IH$_1$ 
& $7.3 $
& $ 4.2 $
& --
& $4.2$
& --
& $6.5$
& $\mathbf{5.5}$ \\

IH$_2$ 
& $5.0$
& $ 5.2$
& --
& $ 6.5$
& --
& --
& $\mathbf{5.6}$ \\

IH$_3$ 
& $7.8$
& $3.2 $
& $4.4$
& $ 6.6 $
& --
& --
& $\mathbf{5.5}$ \\

\bottomrule
\end{tabular}
\end{table}

\subsection{Quantitative agreement and error analysis}
\label{sec:error}
To quantify the reconstruction accuracy across all considered chains, we compute the MAE between the experimental and reconstructed $dI/dV$ maps.
The MAE is defined as
\begin{equation}
\mathrm{MAE_{\text{maps}}}
=
\frac{1}{N}
\sum_{n=1}^{N_n}
\sum_{i=1}^{N_V}
\left|
\left(\frac{dI}{dV}\right)^{\mathrm{pred}}_{n,V_i}
-
\left(\frac{dI}{dV}\right)^{\mathrm{exp}}_{n,V_i}
\right|,
\end{equation}
where $N = N_n N_V$ are the numbers of sites and bias-voltage points, respectively. This metric quantifies the average point-wise difference between the predicted and experimental $dI/dV$ maps, providing a global measure of reconstruction accuracy.
As shown in Tab.~\ref{tab:mae_maps}, the reconstruction yields consistently low errors across all chains, with values on the order of $\sim 5 \%$. Importantly, the MAE remains comparable between Chains IH, indicating that the model robustly captures the inhomogeneous regime, while similarly low errors are obtained for Chain H in the near-uniform case.
These results demonstrate that the Hamiltonian learning framework generalizes across different realizations and geometries, providing a reliable and quantitative description of experimental STM spectroscopy in both homogeneous and inhomogeneous regimes.

\suppsection{Chemical synthesis of the precursor molecule}
\label{sec:precursor}
\subsection{Synthesis of Compound \compound{1}}
\begin{figure}[htbp]
    \centering
    \includegraphics[width=\linewidth]{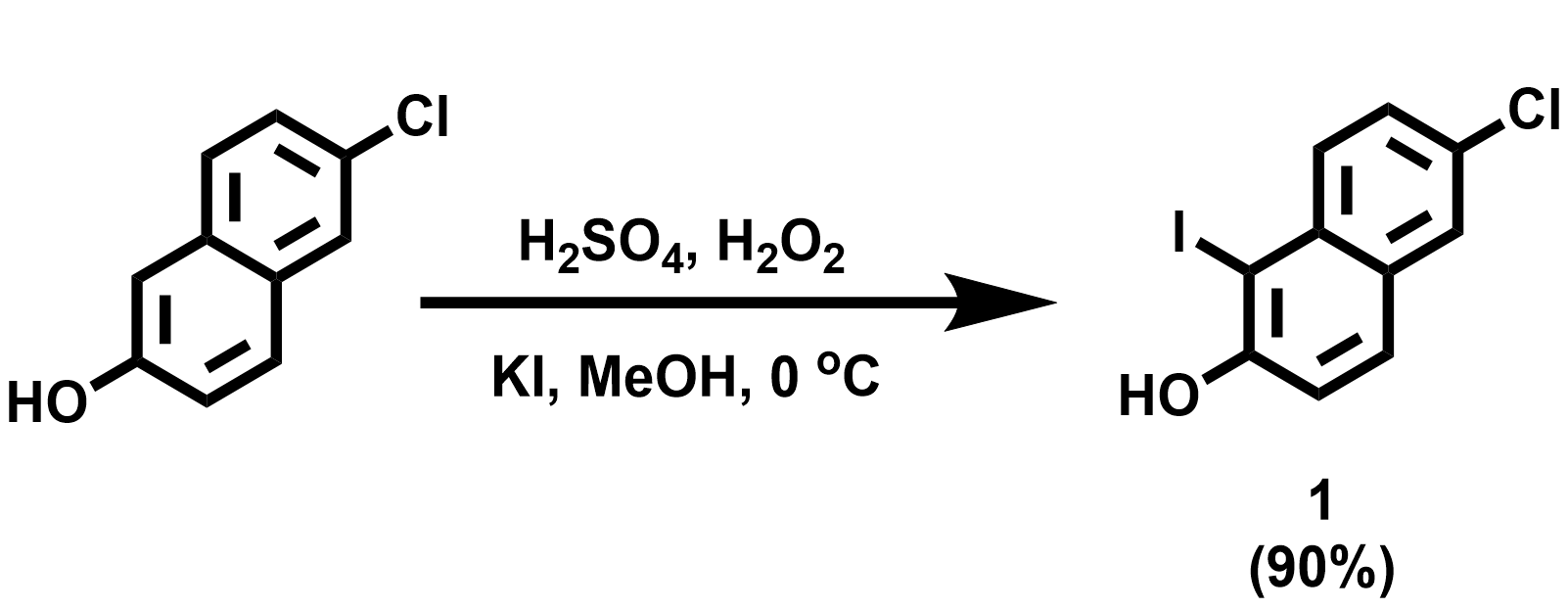}
    \label{fig:c1}
\end{figure} 
In a 100\,mL round glass, sulfuric acid (1.70\,g, 17.31\,mmol) was added to a
solution of 6-chloronaphthalen-2-ol (2.06\,g, 11.54\,mmol) and potassium iodide
(1.92\,g, 11.54\,mmol) in methanol (18\,mL) at 0\,\textdegree C. When the white
precipitate formed, hydrogen peroxide (30\% aqueous solution, 784.85\,mg,
23.07\,mmol) was added. After 1.5\,h, the mixture was filtered, and the filtrate
was concentrated. The residue was dissolved in dichloromethane, washed with
sodium thiosulfate aqueous solution and water. After drying over magnesium
sulfate, the solvent was evaporated \textit{in vacuo} and then purified by silica
gel column chromatography (eluent: iso-hexane/DCM\,=\,2:1) to yield white solid
\compound{1} (3.16\,g, 90\%).

\medskip
\noindent
$^{1}$H\,NMR (300\,MHz, \nmrsolv{CDCl$_3$}): $\delta$\,7.89 (d, $J = 9.0$\,Hz, 1H),
7.75 (d, $J = 2.2$\,Hz, 1H), 7.66 (d, $J = 8.8$\,Hz, 1H),
7.48 (dd, $J = 9.0$, $2.2$\,Hz, 1H), 7.29 (s, 1H), 5.80 (s, 1H).

\medskip
\noindent
$^{13}$C\,NMR (76\,MHz, \nmrsolv{CDCl$_3$}): $\delta$\,154.08, 133.29, 132.06,
130.16, 130.03, 129.75, 128.89, 126.87, 117.56, 85.96.

\medskip
\noindent
HR-MS MALDI-TOF ($m/z$): calculated for \ch{C10H6ClIO} [M$^+$], 303.9152;
found, 303.9151; error $= -0.13$\,ppm.

\medskip
\noindent
\begin{figure}[htbp]
    \centering
    \includegraphics[width=\linewidth]{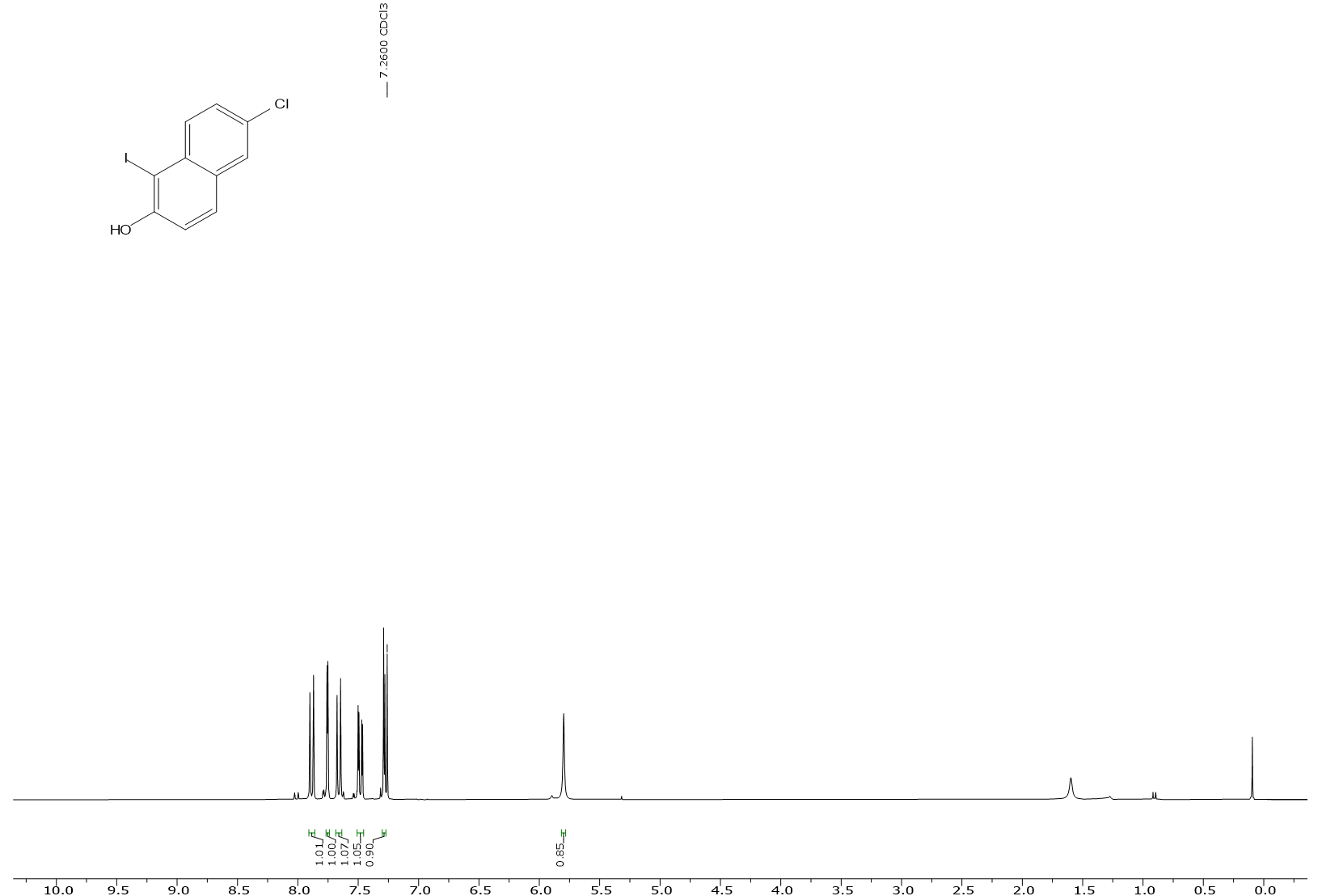}
    \caption{ $^{1}$H\,NMR spectrum (300\,MHz, CDCl$_3$) of \compound{1}.}
    \label{fig:1}
\end{figure}

\medskip
\noindent
\begin{figure}[htbp]
    \centering
    \includegraphics[width=\linewidth]{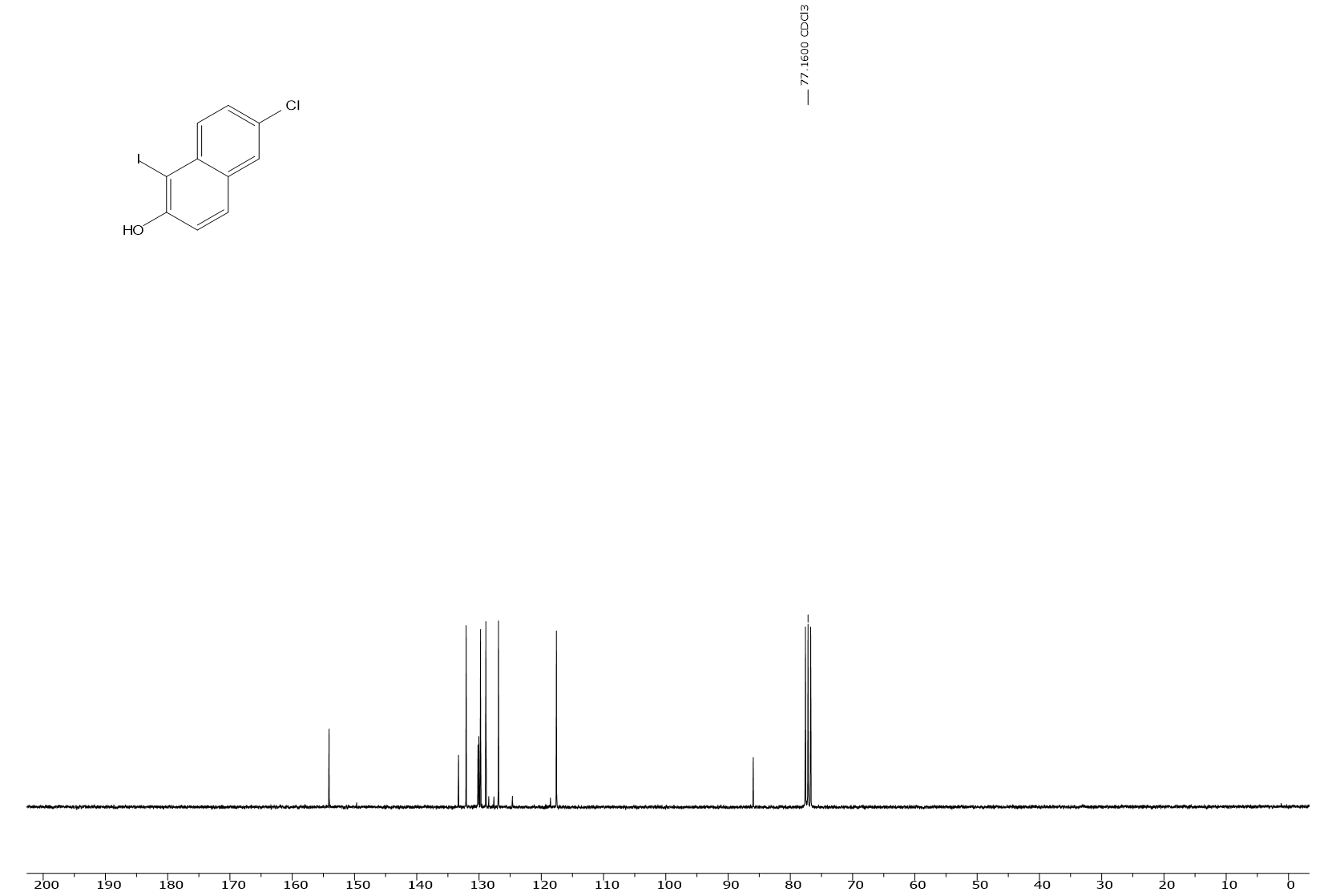}
    \caption{$^{13}$C\,NMR (76\,MHz, CDCl$_3$) of \compound{1}}
    \label{fig:2}
\end{figure}

\subsection{Synthesis of Compound \compound{2}}
\begin{figure}[htbp]
    \centering
    \includegraphics[width=\linewidth]{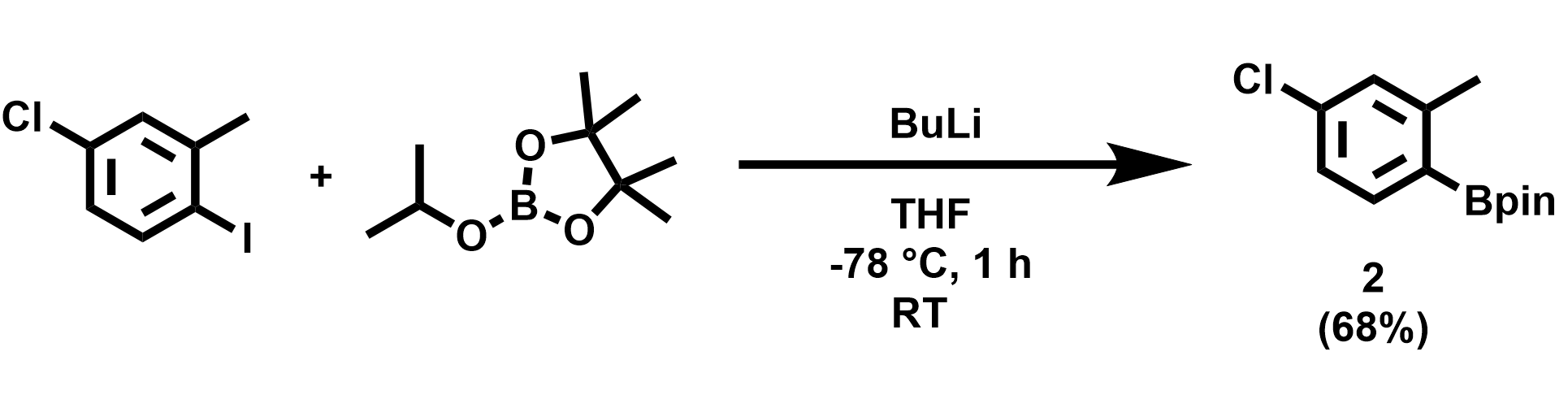}
    \label{fig:c2}
\end{figure} 
To a solution of 4-chloro-1-iodo-2-methylbenzene (3.90\,g, 15.00\,mmol) in
anhydrous tetrahydrofuran (30\,mL) at $-78$\,\textdegree C under an argon
atmosphere, \textit{n}-butyllithium (1.6\,M in hexanes, 9.2\,mL, 14.70\,mmol)
was added dropwise. The mixture was stirred at $-78$\,\textdegree C for 1\,h,
forming a white suspension. 2-isopropoxy-4,4,5,5-tetramethyl-1,3,2-dioxaborolane
(3.77\,g, 20.00\,mmol) was then added to the reaction mixture at
$-78$\,\textdegree C. The cooling bath was removed, and the solution was allowed
to warm slowly to room temperature and stirred overnight. After completion of the
reaction (monitored by TLC), the mixture was quenched with water, and the aqueous
layer was extracted with ethyl acetate ($3 \times 30$\,mL). The combined organic
layers were washed with brine ($2 \times 20$\,mL), dried over anhydrous
\ch{MgSO4}, filtered, and concentrated under reduced pressure. The crude product
was purified by flash column chromatography on silica gel (eluent:
iso-hexane/DCM\,=\,$1:0 \to 1:1$), to afford the desired boronate ester as a
white solid (2.67\,g, 68\%).

\medskip
\noindent
$^{1}$H\,NMR (300\,MHz, \nmrsolv{CDCl$_3$}): $\delta$\,7.68 (d, $J = 7.8$\,Hz,
1H), 7.18--7.11 (m, 2H), 2.51 (s, 3H), 1.34 (s, 12H).

\medskip
\noindent
$^{13}$C\,NMR (76\,MHz, \nmrsolv{CDCl$_3$}): $\delta$\,147.03, 137.38, 136.92,
129.89, 125.07, 83.75, 25.03, 22.14.

\medskip
\noindent
HR-MS MALDI-TOF ($m/z$): calculated for \ch{C13H18BClO2} [M$+$H]$^+$,
251.1125; found, 251.1129; error $= +1.68$\,ppm.

\medskip
\noindent
\begin{figure}[htbp]
    \centering
    \includegraphics[width=\linewidth]{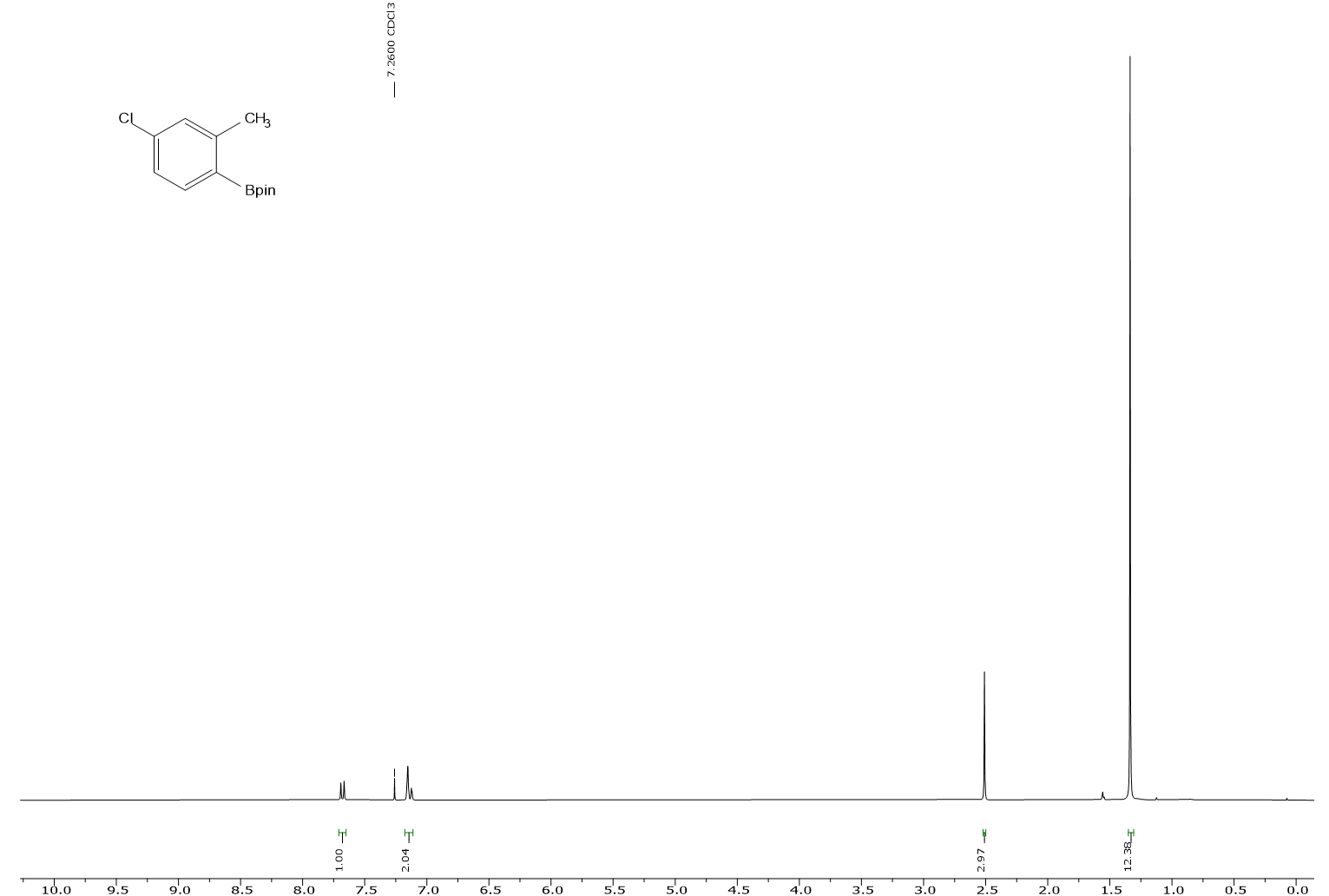}
    \caption{ $^{1}$H\,NMR (300\,MHz, CDCl$_3$) of \compound{2}.}
    \label{fig:3}
\end{figure}

\medskip
\noindent
\begin{figure}[htbp]
    \centering
    \includegraphics[width=\linewidth]{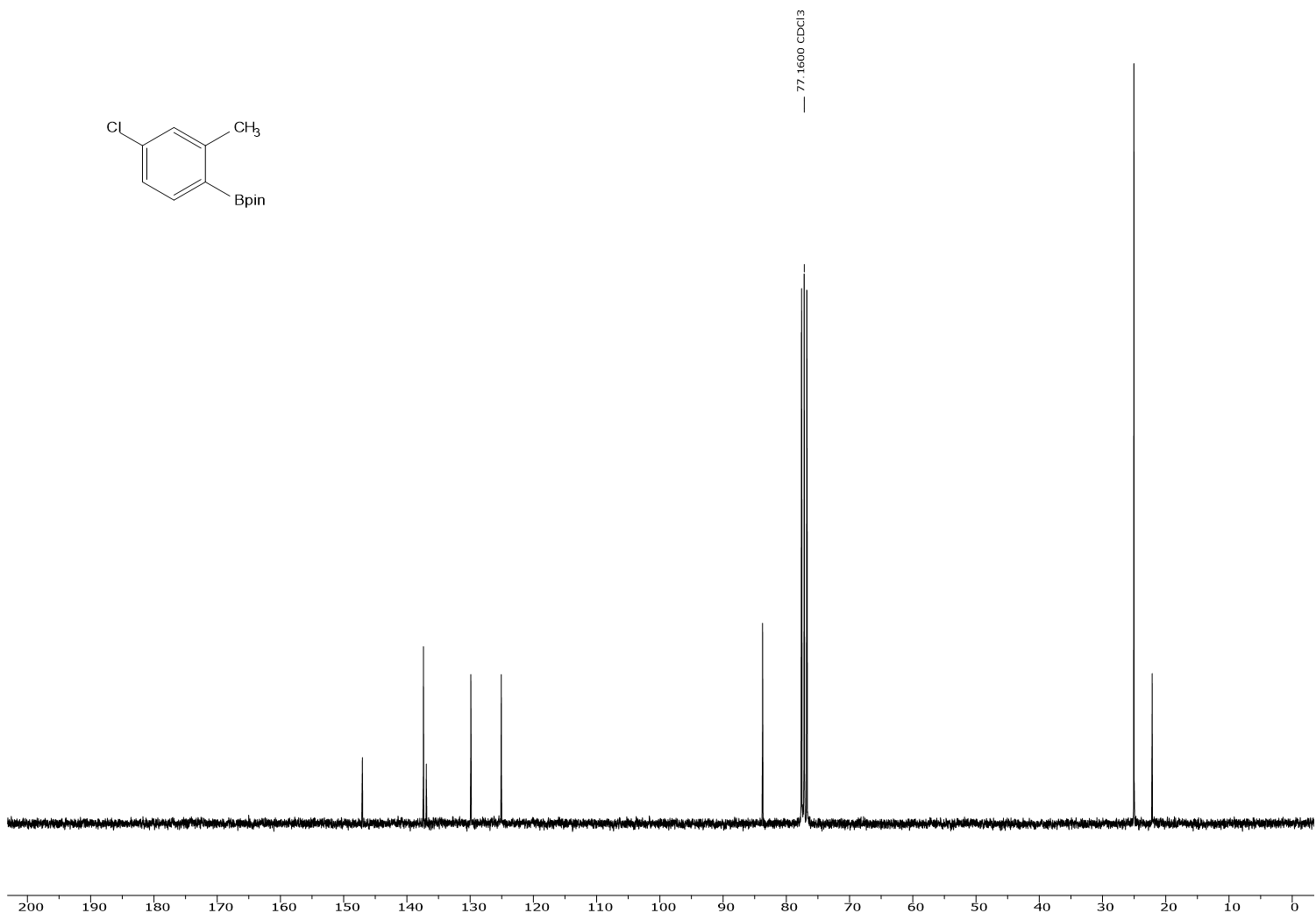}
    \caption{$^{13}$C\,NMR (76\,MHz, CDCl$_3$) of \compound{2}.}
    \label{fig:4}
\end{figure}

\subsection{Synthesis of Compound \compound{3}}
\begin{figure}[htbp]
    \centering
    \includegraphics[width=\linewidth]{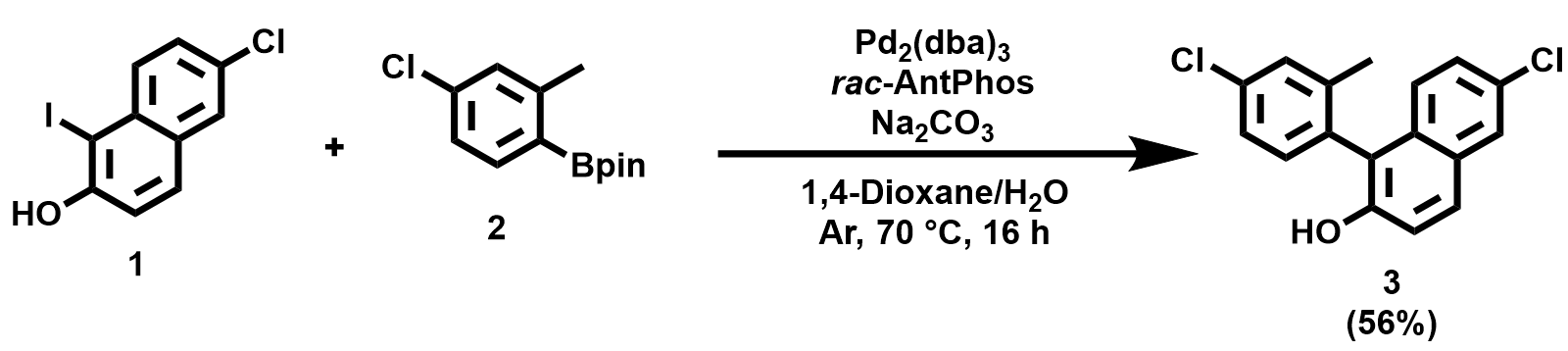}
    \label{fig:c3}
\end{figure} 
A 25\,mL dried Schlenk flask was charged with compound \compound{2} (586\,mg,
2.32\,mmol), compound \compound{1} (680\,mg, 2.24\,mmol), sodium carbonate
(760\,mg, 7.17\,mmol), tris(dibenzylidene-acetone)dipalladium(0) (220\,mg,
0.24\,mmol), and \textit{rac}-AntPhos (176\,mg, 0.47\,mmol). The system was
purged with argon for 10\,min. Degassed 1,4-dioxane (15\,mL) and degassed water
(3\,mL) were added via syringe, and the resulting red-brown suspension was stirred
vigorously in a 70\,\textdegree C oil bath overnight. After completion of the
reaction (monitored by TLC), the mixture was cooled to room temperature, quenched
with saturated aqueous ammonium chloride solution, and extracted with
dichloromethane ($3 \times 20$\,mL). The combined organic layers were dried over
anhydrous magnesium sulfate, filtered, and concentrated under reduced pressure.
The crude residue was purified by silica gel column chromatography (eluent:
iso-hexane/DCM\,=\,$4:1 \to 2:1 \to 1:1$) to afford the desired biaryl product as
a yellow oily semi-solid (396\,mg, 56\%).

\medskip
\noindent
$^{1}$H\,NMR (300\,MHz, \nmrsolv{CD$_2$Cl$_2$}): $\delta$\,7.82 (d, $J = 2.2$\,Hz,
1H), 7.76 (d, $J = 8.9$\,Hz, 1H), 7.46 (d, $J = 2.2$\,Hz, 1H), 7.38 (dd,
$J = 8.1$, $2.3$\,Hz, 1H), 7.30 (d, $J = 3.5$\,Hz, 1H), 7.28--7.25 (m, 1H),
7.20 (d, $J = 8.1$\,Hz, 1H), 7.11 (d, $J = 9.0$\,Hz, 1H), 4.95 (s, 1H),
1.98 (s, 3H).

\medskip
\noindent
$^{13}$C\,NMR (76\,MHz, \nmrsolv{CD$_2$Cl$_2$}): $\delta$\,150.92, 141.38,
135.10, 133.19, 131.79, 131.78, 131.33, 129.96, 129.45, 129.21, 127.74, 127.43,
127.12, 126.35, 119.71, 118.98, 19.61.

\medskip
\noindent
HR-MS MALDI-TOF ($m/z$): calculated for \ch{C17H12Cl2O} [M$-$H]$^-$, 302.0265;
found, 302.0263; error $= -0.59$\,ppm.
\medskip
\noindent
\begin{figure}[htbp]
    \centering
    \includegraphics[width=\linewidth]{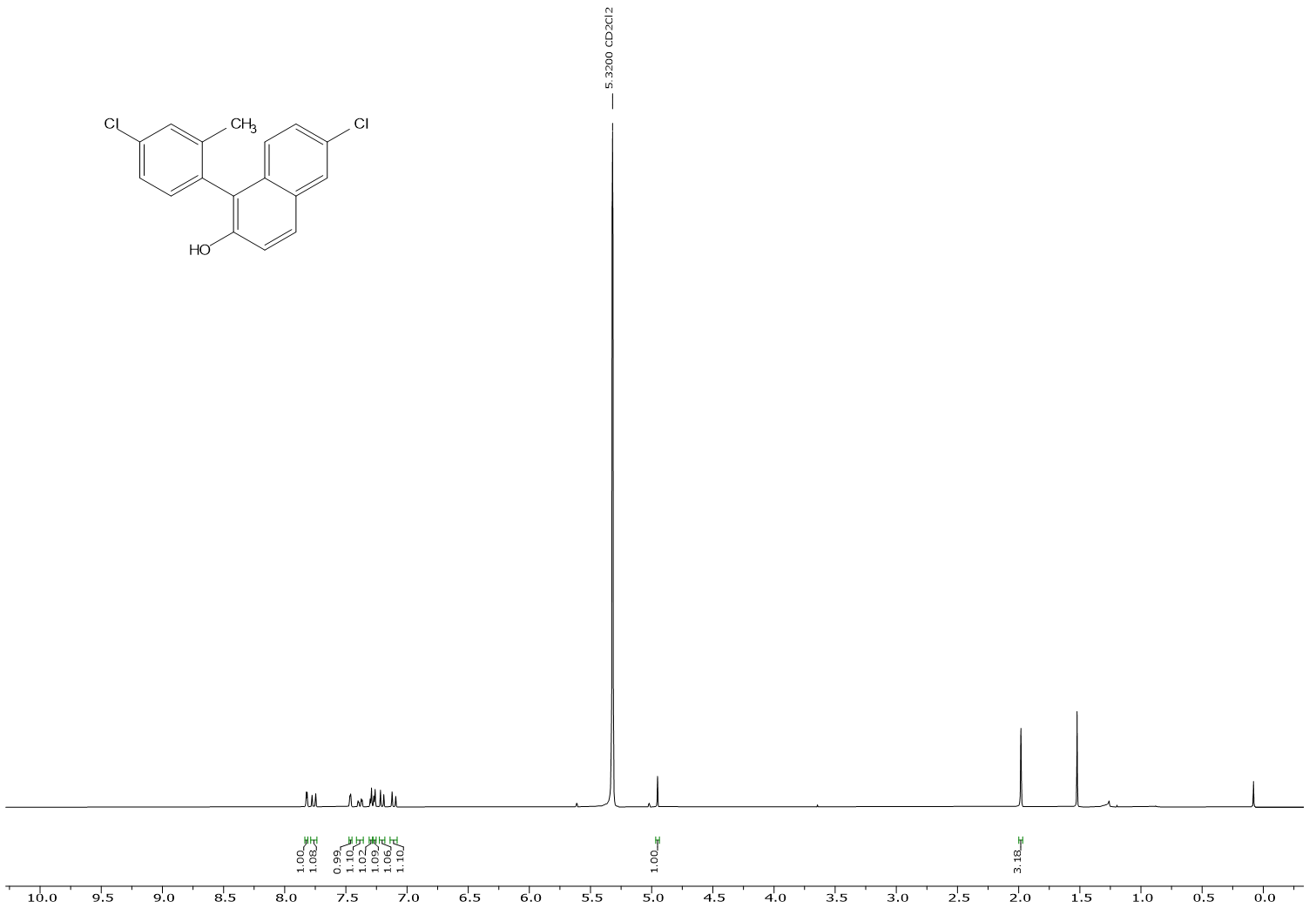}
    \caption{$^{1}$H\,NMR (300\,MHz, CD$_2$Cl$_2$) of \compound{3}.}
    \label{fig:5}
\end{figure}

\medskip
\noindent
\begin{figure}[htbp]
    \centering
    \includegraphics[width=\linewidth]{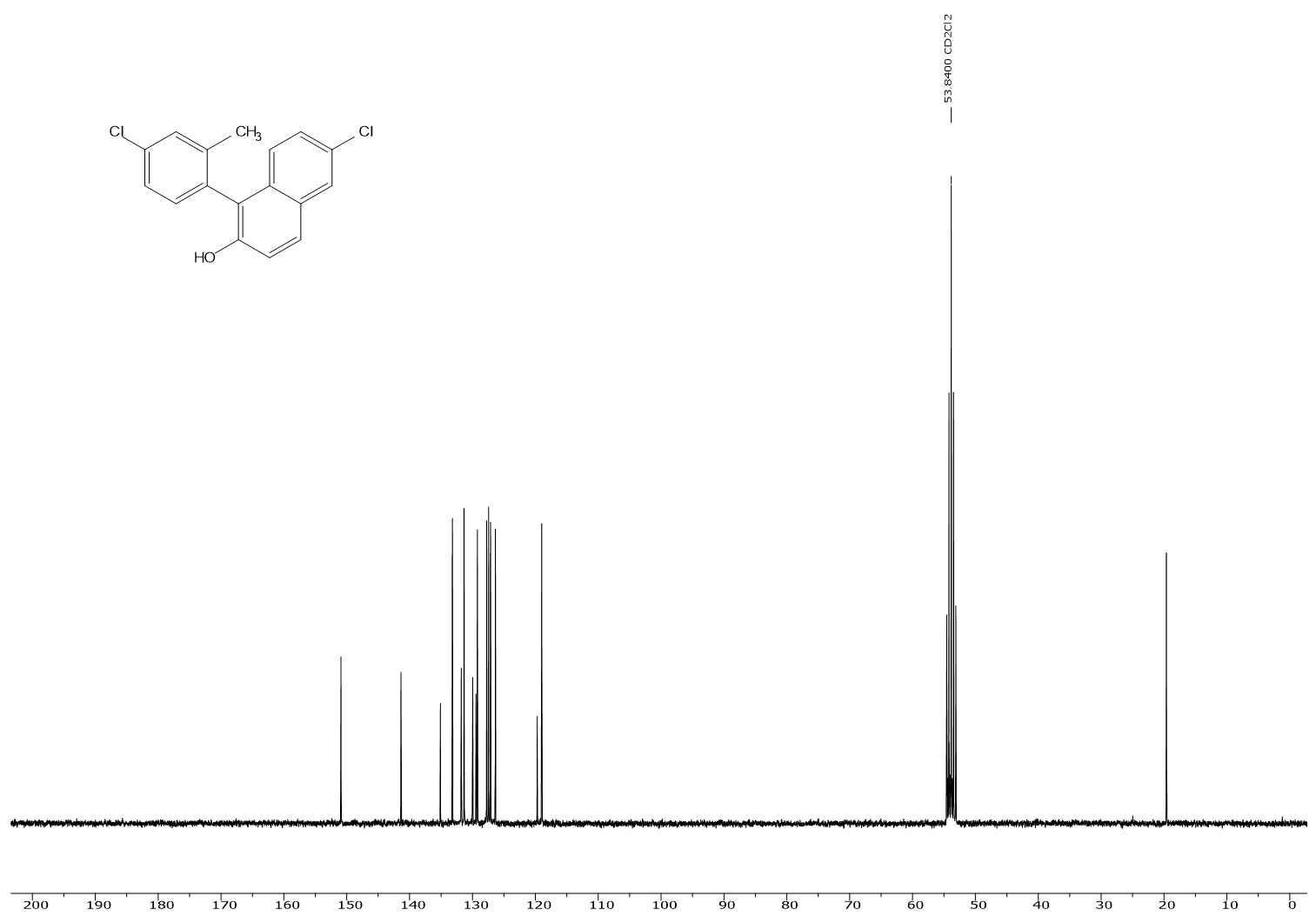}
    \caption{$^{13}$C\,NMR (76\,MHz, CD$_2$Cl$_2$) of \compound{3}.}
    \label{fig:6}
\end{figure}

\subsection{Synthesis of Compound \compound{4}}
\begin{figure}[htbp]
    \centering
    \includegraphics[width=\linewidth]{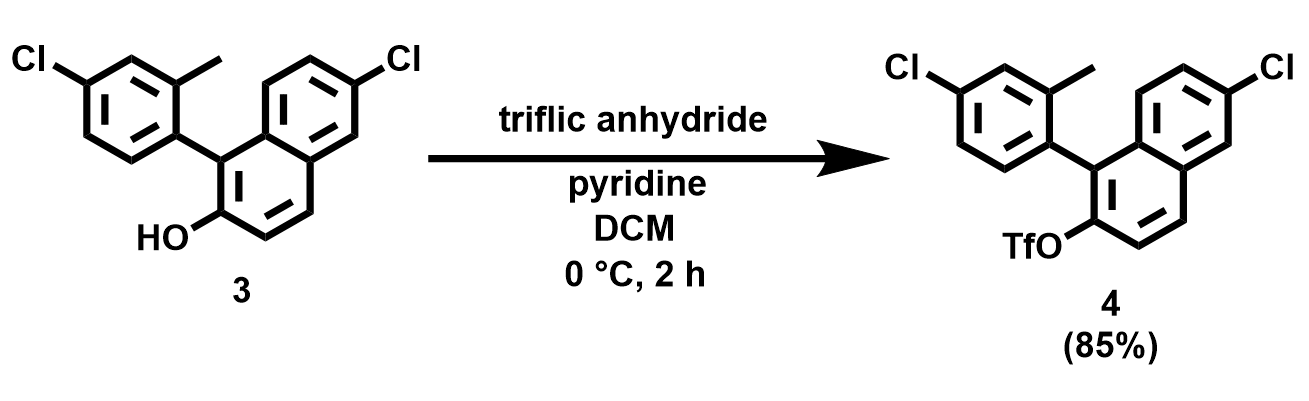}
    \label{fig:c4}
\end{figure} 
To a solution of 1-(4-chloro-2-methylphenyl)-6-chloronaphthalen-2-ol
(370.00\,mg, 1.22\,mmol) and pyridine (193.07\,mg, 2.44\,mmol) in
dichloromethane (2.48\,mL) was added triflic anhydride (413.17\,mg, 1.46\,mmol)
dropwise at 0\,\textdegree C under air. The mixture was allowed to warm to room
temperature and stirred for 2\,h. After completion of the reaction (monitored by
TLC), the mixture was quenched with 1\,M HCl and extracted with dichloromethane.
The combined organic layers were dried over anhydrous magnesium sulfate, filtered,
and concentrated under reduced pressure. The crude residue was purified by silica
gel column chromatography (eluent: isohexane/DCM\,=\,4:1) to afford the desired
triflate product as a yellow gel (451.49\,mg, 85\%).

\medskip
\noindent
$^{1}$H\,NMR (300\,MHz, \nmrsolv{CDCl$_3$}): $\delta$\,7.94 (d, $J = 2.1$\,Hz,
1H), 7.89 (d, $J = 9.1$\,Hz, 1H), 7.51 (d, $J = 9.1$\,Hz, 1H), 7.45--7.32
(m, 4H), 7.17 (d, $J = 8.2$\,Hz, 1H), 1.97 (s, 3H).

\medskip
\noindent
$^{13}$C\,NMR (76\,MHz, \nmrsolv{CDCl$_3$}): $\delta$\,144.32, 139.61, 135.09,
133.47, 133.30, 132.27, 131.34, 131.27, 130.60, 130.54, 129.59, 128.87, 128.12,
127.14, 126.34, 120.85, 118.44 (q, $J = 322$\,Hz), 19.74.

\medskip
\noindent
$^{19}$F\,NMR (283\,MHz, \nmrsolv{CDCl$_3$}): $\delta$\,$-74.28$.

\medskip
\noindent
HR-MS MALDI-TOF ($m/z$): calculated for \ch{C18H11Cl2F3O3S} [M]$^+$, 433.9758;
found, 433.9765; error $= +1.54$\,ppm.
\medskip
\noindent
\begin{figure}[htbp]
    \centering
    \includegraphics[width=\linewidth]{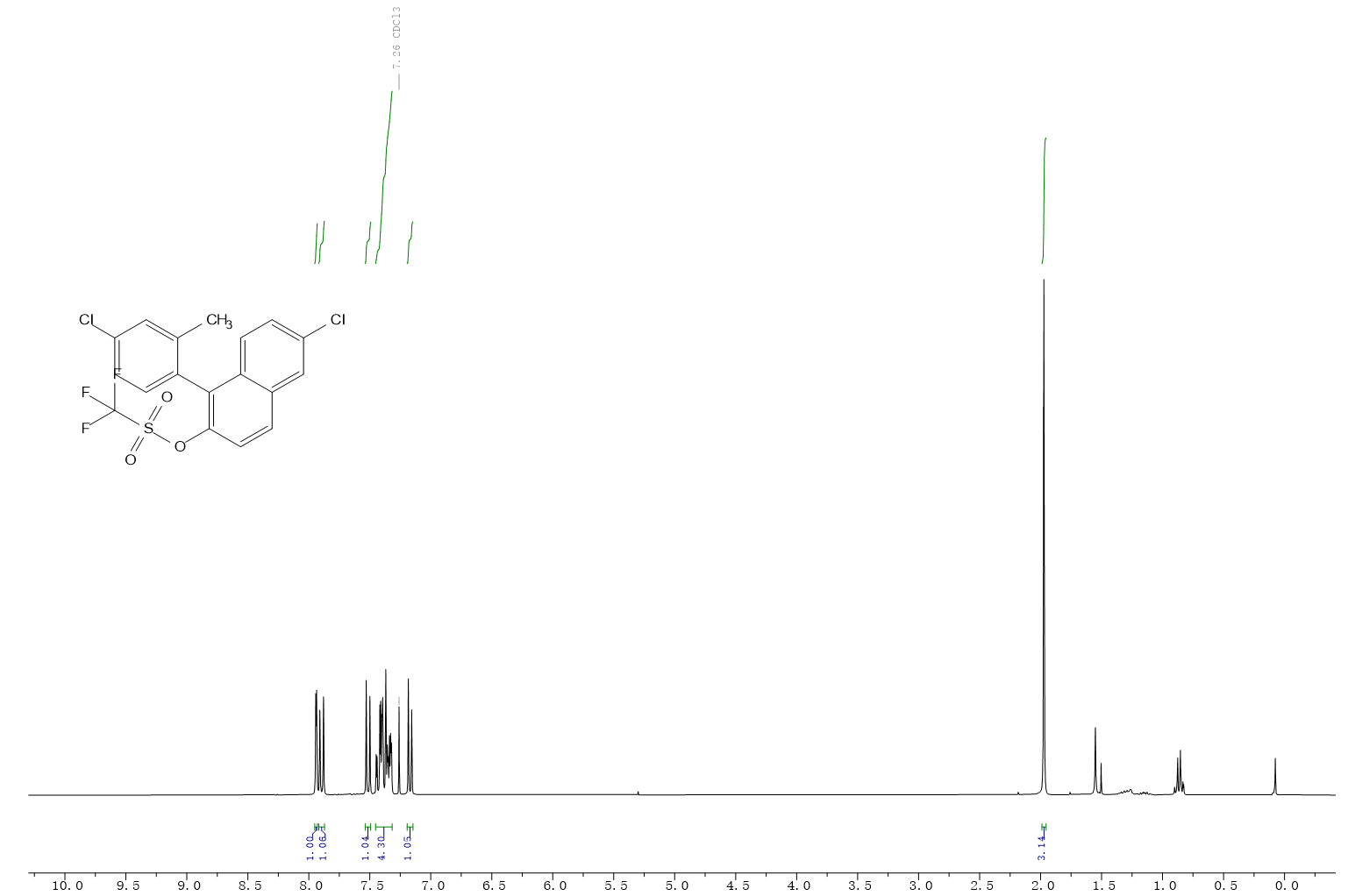}
    \caption{$^{1}$H\,NMR (300\,MHz, CDCl$_3$) of \compound{4}.}
    \label{fig:7}
\end{figure}
\medskip
\noindent
\begin{figure}[htbp]
    \centering
    \includegraphics[width=\linewidth]{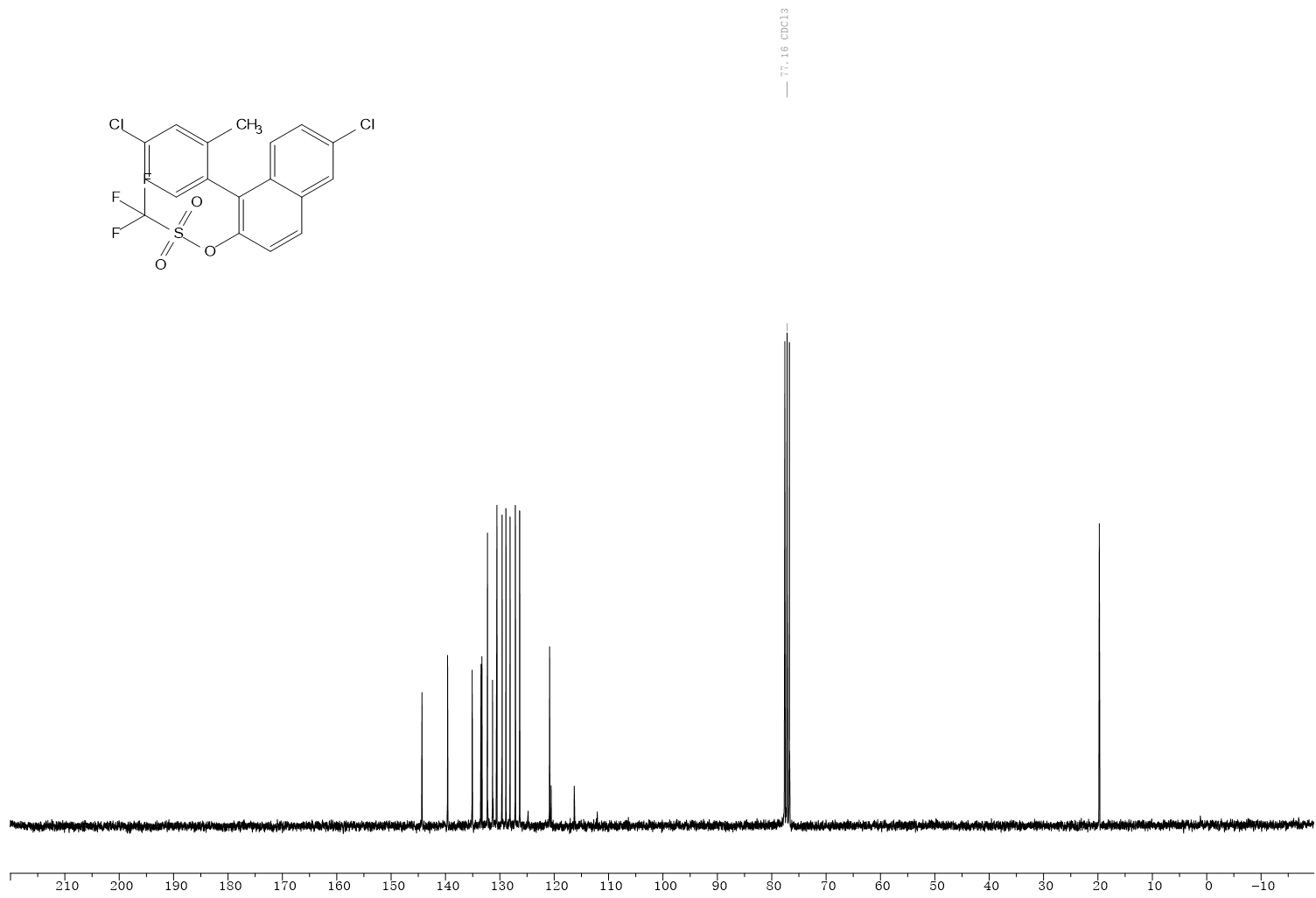}
    \caption{$^{13}$C\,NMR (76\,MHz, CDCl$_3$) of \compound{4}.}
    \label{fig:8}
\end{figure}
\medskip
\noindent
\begin{figure}[htbp]
    \centering
    \includegraphics[width=\linewidth]{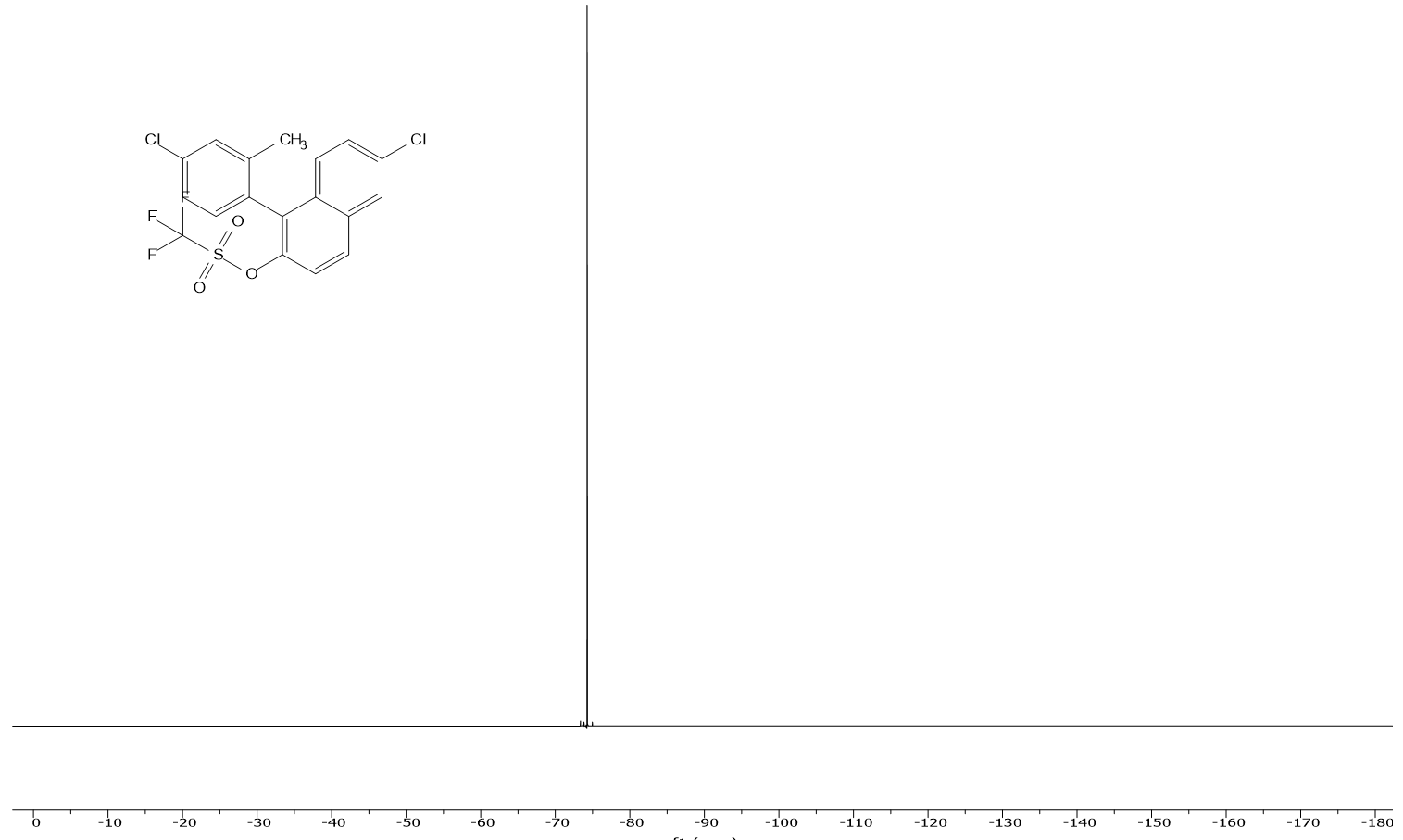}
    \caption{$^{19}$F\,NMR (283\,MHz, CDCl$_3$) of \compound{4}.}
    \label{fig:9}
\end{figure}

\subsection{Synthesis of Compound \compound{5}}
\begin{figure}[htbp]
    \centering
    \includegraphics[width=\linewidth]{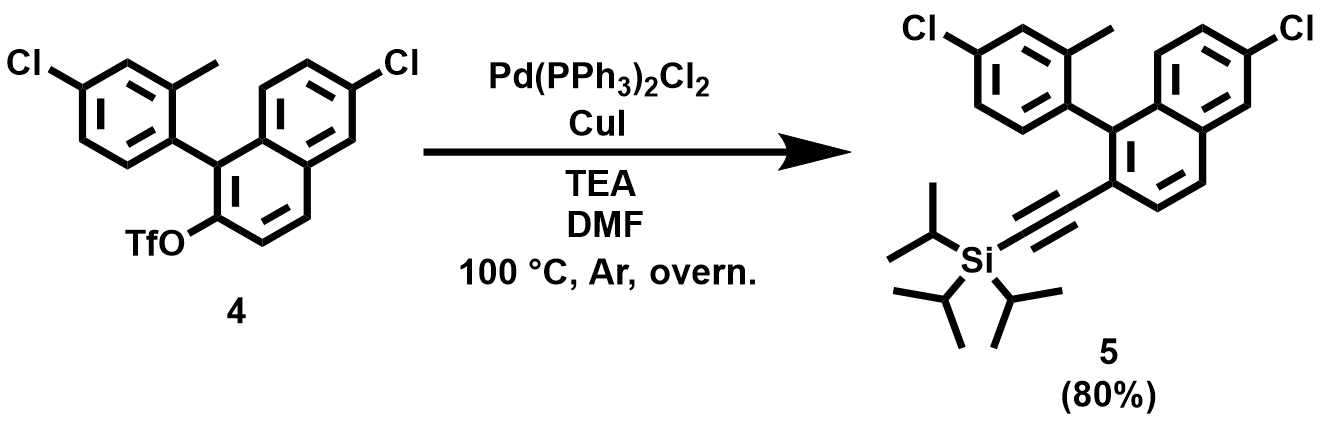}
    \label{fig:c5}
\end{figure} 
To a degassed suspension of compound \compound{4} (366.60\,mg, 842.30\,\textmu mol)
and CuI (64.17\,mg, 336.92\,\textmu mol) in a mixture of triethylamine (0.12\,mL)
and DMF (21.0\,mL) were added \ch{PdCl2(PPh3)2} (118.24\,mg, 168.46\,\textmu mol)
and triisopropylsilylacetylene (921.72\,mg, 1.13\,mL, 5.05\,mmol) under an argon
atmosphere. The reaction mixture was stirred overnight at 100\,\textdegree C,
cooled to room temperature, diluted in dichloromethane, and washed with saturated
aqueous ammonium chloride solution. The organic layer was collected, dried over
magnesium sulfate, and evaporated. The crude product was purified by silica gel
column chromatography (eluent: iso-hexane/DCM\,=\,4:1) to afford compound
\compound{5} as a colourless oil (315.06\,mg, 80\%).

\medskip
\noindent
$^{1}$H\,NMR (300\,MHz, \nmrsolv{CD$_2$Cl$_2$}): $\delta$\,7.87 (d, $J = 2.0$\,Hz,
1H), 7.75 (d, $J = 8.6$\,Hz, 1H), 7.65 (d, $J = 8.5$\,Hz, 1H), 7.33 (dd,
$J = 6.2$, $2.2$\,Hz, 2H), 7.30--7.26 (m, 2H), 7.11 (d, $J = 8.1$\,Hz, 1H),
1.97 (s, 3H), 0.96 (s, 21H).

\medskip
\noindent
$^{13}$C\,NMR (76\,MHz, \nmrsolv{CD$_2$Cl$_2$}): $\delta$\,142.29, 139.54,
137.22, 134.07, 133.93, 132.81, 131.83, 130.69, 130.37, 130.32, 128.26, 127.90,
127.16, 127.10, 126.35, 121.54, 105.96, 95.94, 19.70, 18.64, 11.57.
\medskip
\noindent
\begin{figure}[htbp]
    \centering
    \includegraphics[width=\linewidth]{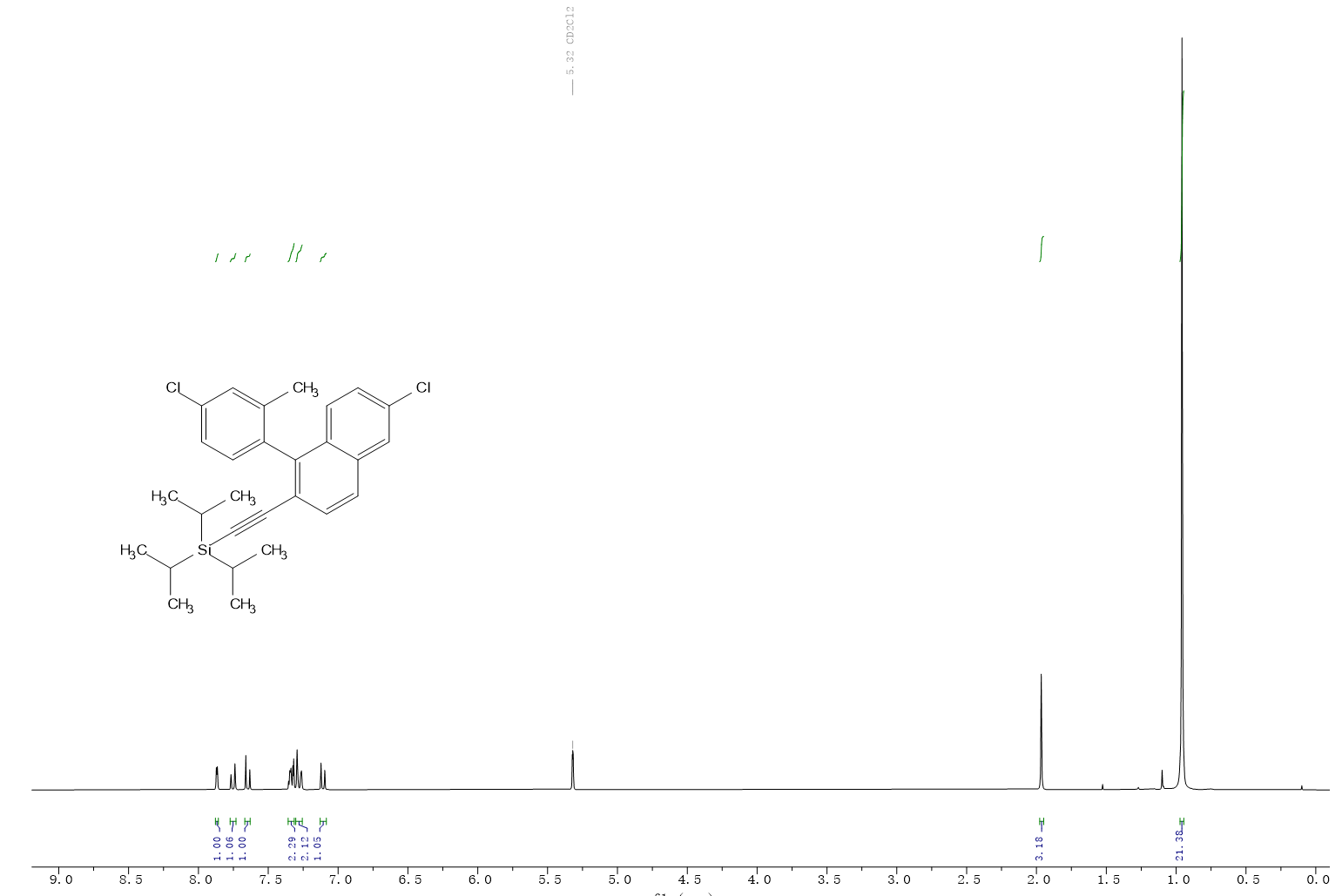}
    \caption{$^{1}$H\,NMR (300\,MHz, CD$_2$Cl$_2$) of \compound{5}.}
    \label{fig:10}
\end{figure}
\medskip
\noindent
\begin{figure}[htbp]
    \centering
    \includegraphics[width=\linewidth]{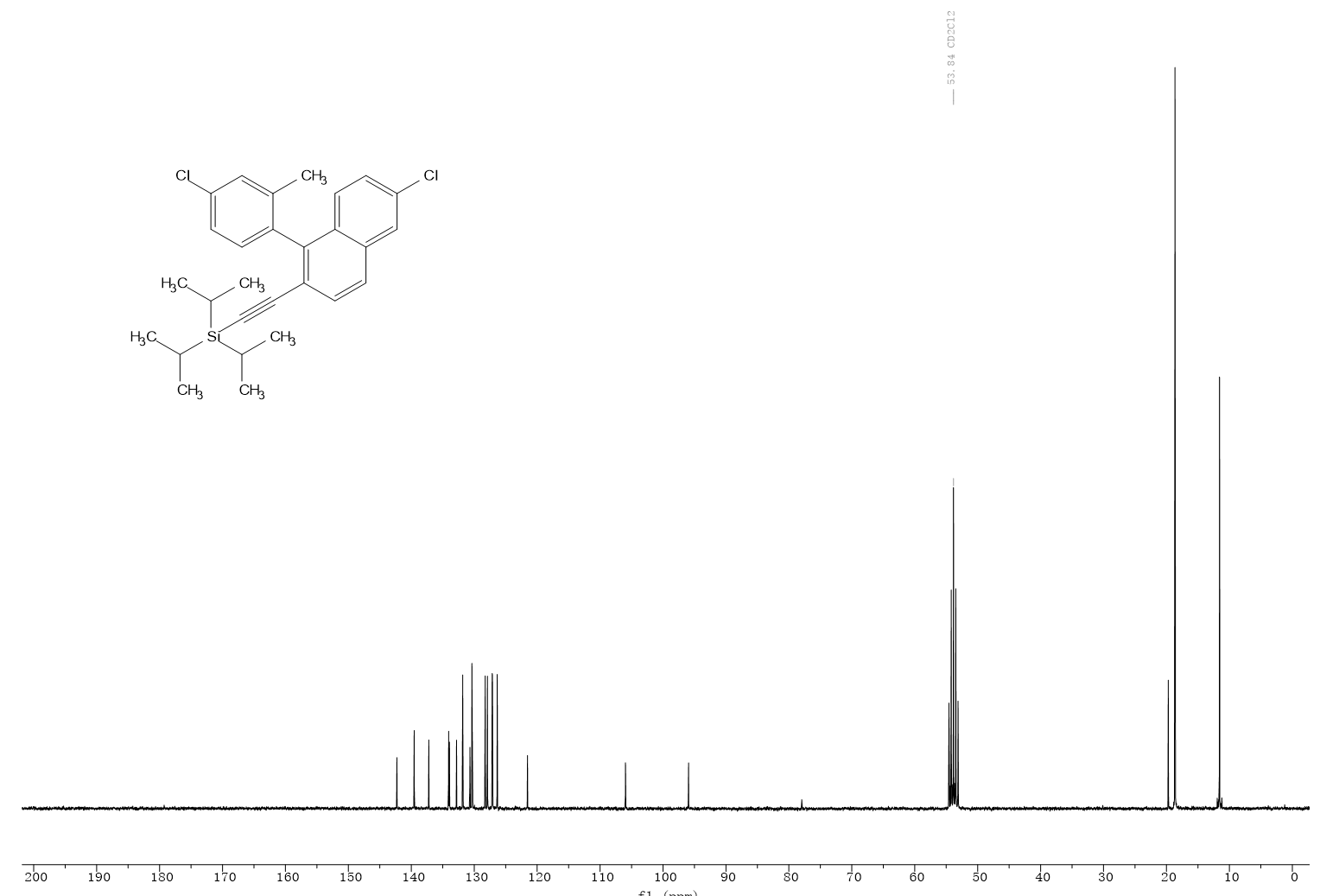}
    \caption{$^{13}$C\,NMR (76\,MHz, CD$_2$Cl$_2$) of \compound{5}.}
    \label{fig:11}
\end{figure}

\subsection{Synthesis of Compound \compound{6}}
\begin{figure}[htbp]
    \centering
    \includegraphics[width=\linewidth]{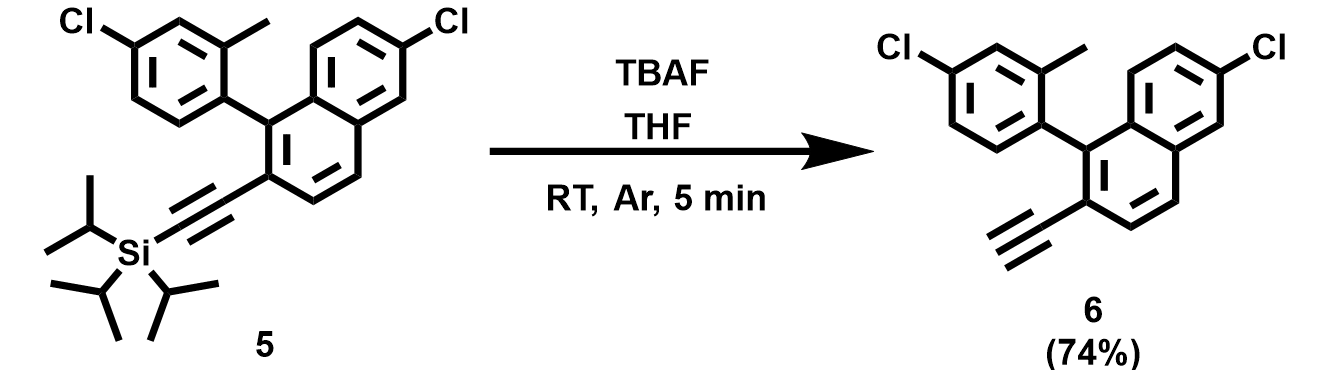}
    \label{fig:c6}
\end{figure} 

Argon was bubbled through a solution of compound \compound{5} (14.40\,mg,
30.80\,\textmu mol) in THF (0.38\,mL) for 5\,min, and then TBAF (1\,M in THF,
46.20\,\textmu L, 46.20\,\textmu mol) was added dropwise. After 5\,min, the
reaction was quenched with methanol. The solvent was removed under vacuum. The
residue was purified by silica gel column chromatography (eluent:
iso-hexane/DCM\,=\,4:1) to afford compound \compound{6} (7.09\,mg, 74\%).

\medskip
\noindent
$^{1}$H\,NMR (300\,MHz, \nmrsolv{CD$_2$Cl$_2$}): $\delta$\,7.89 (d, $J = 2.2$\,Hz,
1H), 7.78 (d, $J = 8.5$\,Hz, 1H), 7.66 (d, $J = 8.6$\,Hz, 1H), 7.40--7.30
(m, 3H), 7.26 (d, $J = 9.0$\,Hz, 1H), 7.13 (d, $J = 8.1$\,Hz, 1H), 3.08
(s, 1H), 1.96 (s, 3H).

\medskip
\noindent
$^{13}$C\,NMR (76\,MHz, \nmrsolv{CD$_2$Cl$_2$}): $\delta$\,142.42, 139.57,
136.77, 134.29, 133.99, 133.11, 131.80, 130.60, 130.42, 130.32, 128.31, 128.05,
127.33, 127.18, 126.27, 120.04, 82.79, 81.58, 19.65.

\medskip
\noindent
HR-MS MALDI-TOF ($m/z$): calculated for \ch{C19H12Cl2} [M$+$H]$^+$, 310.0316;
found, 310.0314; error $= -0.62$\,ppm.
\medskip
\noindent
\begin{figure}[htbp]
    \centering
    \includegraphics[width=\linewidth]{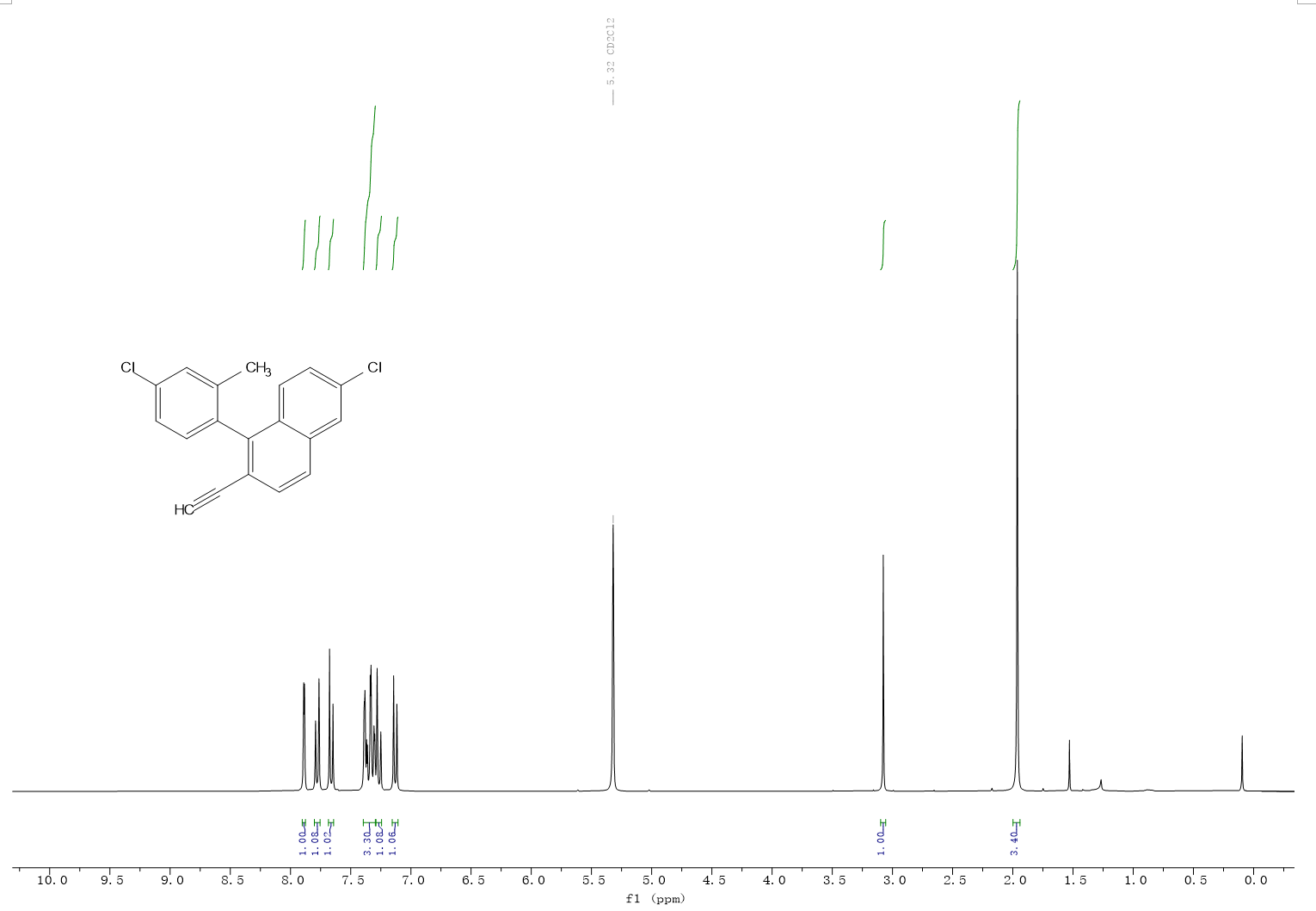}
    \caption{$^{1}$H\,NMR (300\,MHz, CD$_2$Cl$_2$) of \compound{6}.}
    \label{fig:12}
\end{figure}
\medskip
\noindent
\begin{figure}[htbp]
    \centering
    \includegraphics[width=\linewidth]{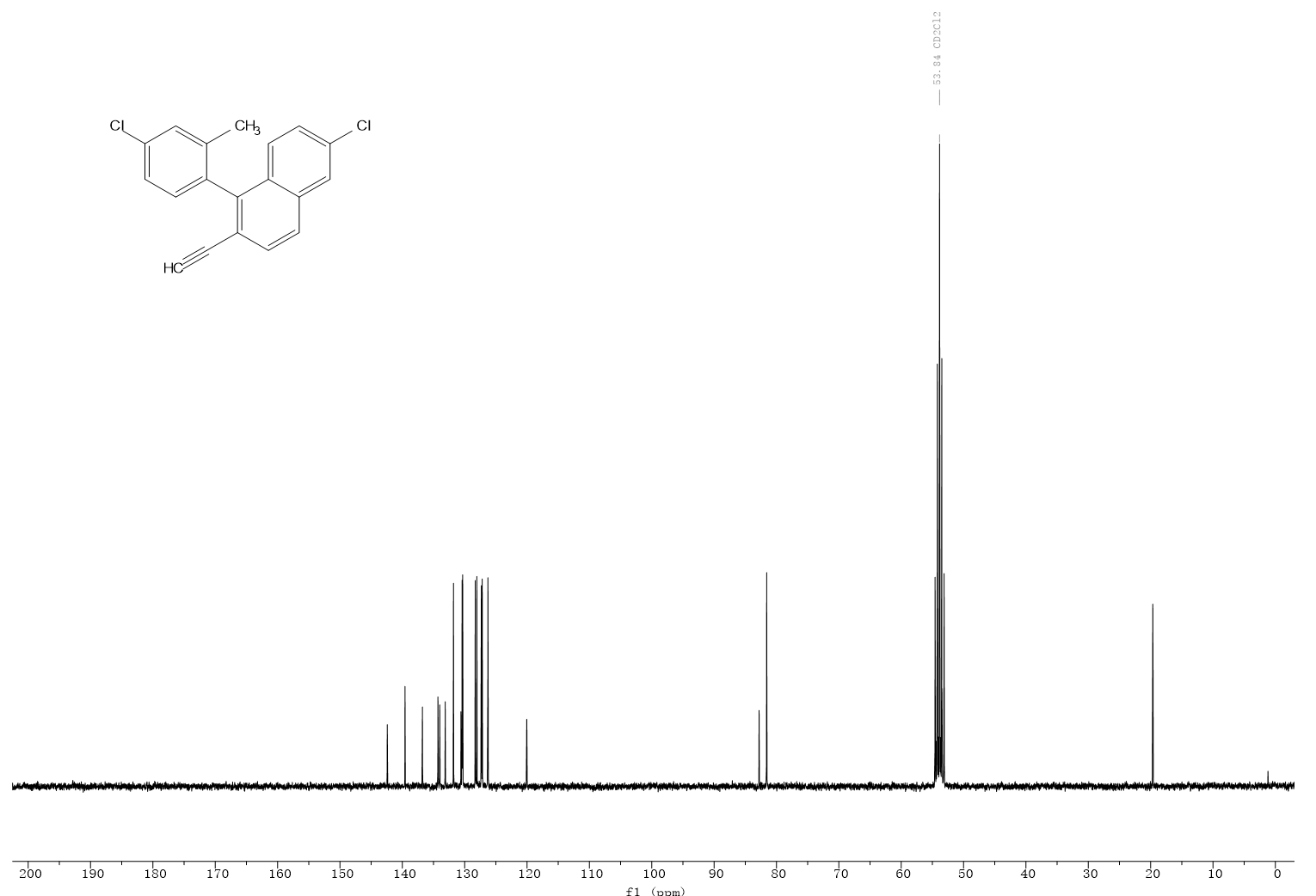}
    \caption{$^{13}$C\,NMR (76\,MHz, CD$_2$Cl$_2$) of \compound{6}.}
    \label{fig:13}
\end{figure}

\subsection{Synthesis of Compound \compound{7}}
\begin{figure}[htbp]
    \centering
    \includegraphics[width=\linewidth]{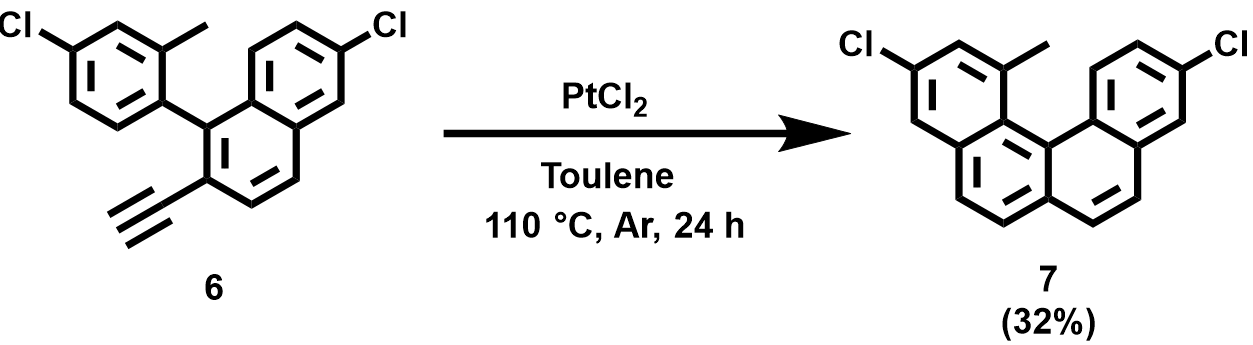}
    \label{fig:c7}
\end{figure} 

Compound \compound{6} (169.00\,mg, 543.05\,\textmu mol) and \ch{PtCl2}
(28.89\,mg, 108.61\,\textmu mol) were suspended in anhydrous toluene (20.00\,mL)
in a Schlenk flask under an argon atmosphere. The reaction mixture was stirred at
110\,\textdegree C for 24\,h. After cooling to room temperature, the mixture was
filtered to remove insoluble residues, and the solvent was removed under reduced
pressure. The crude product was purified by silica gel column chromatography
(eluent: iso-hexane/DCM\,=\,4:1) to afford compound \compound{7} (54.08\,mg,
32\%) as a solid.

\medskip
\noindent
$^{1}$H\,NMR (300\,MHz, \nmrsolv{CDCl$_3$}): $\delta$\,7.96 (dd, $J = 2.2$,
$0.4$\,Hz, 1H), 7.91--7.85 (m, 3H), 7.82 (dd, $J = 2.3$, $0.6$\,Hz, 1H),
7.75 (s, 2H), 7.53--7.48 (m, 2H), 2.37 (s, 3H).

\medskip
\noindent
$^{13}$C\,NMR (76\,MHz, \nmrsolv{CDCl$_3$}): $\delta$\,137.69, 135.11, 133.23,
131.83, 131.74, 131.67, 131.51, 129.92, 128.47, 128.01, 127.33, 127.20, 127.07,
126.92, 126.72, 126.47, 125.79, 124.55, 24.72.

\medskip
\noindent
HR-MS MALDI-TOF ($m/z$): calculated for \ch{C19H12Cl2} [M]$^+$, 310.0316;
found, 310.0323; error $= +2.22$\,ppm.
\medskip
\noindent
\begin{figure}[htbp]
    \centering
    \includegraphics[width=\linewidth]{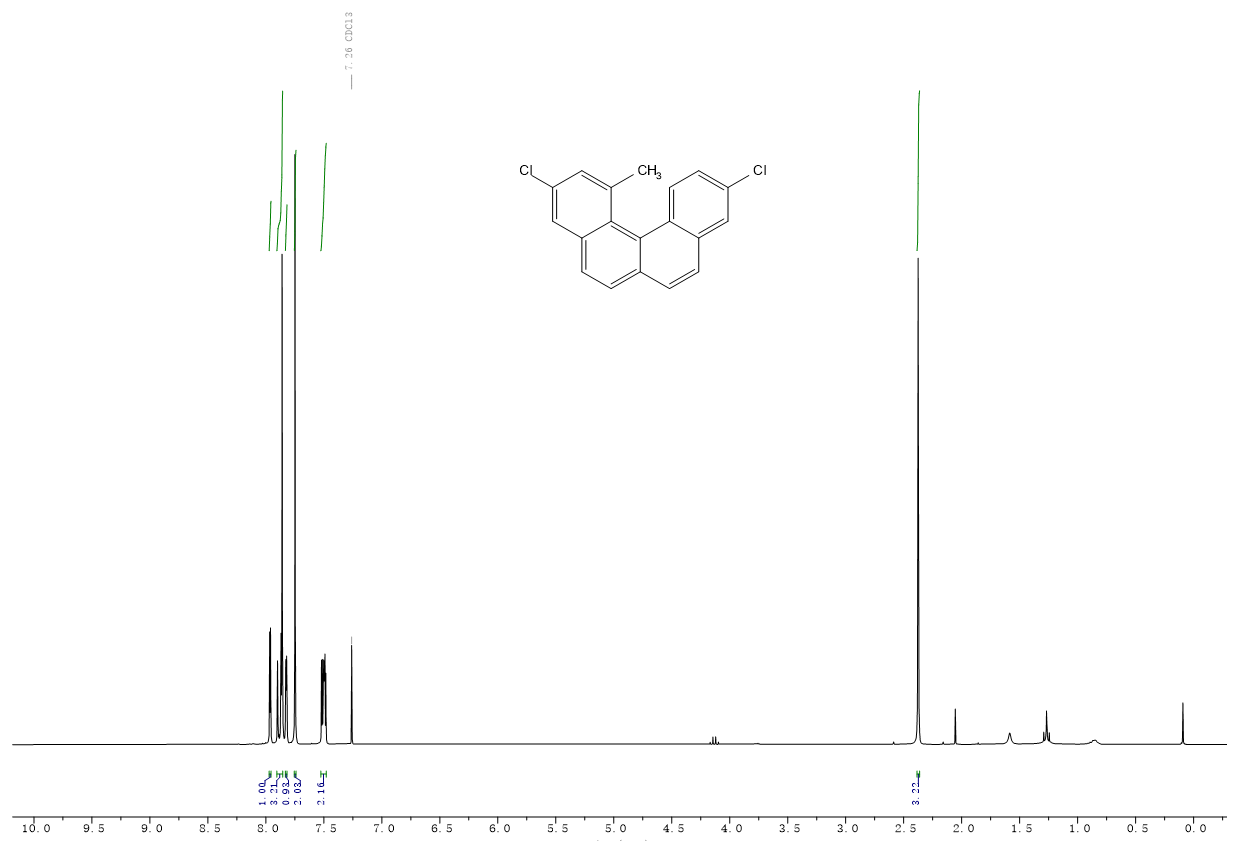}
    \caption{$^{1}$H\,NMR (300\,MHz, CDCl$_3$) of \compound{7}.}
    \label{fig:14}
\end{figure}
\medskip
\noindent
\begin{figure}[htbp]
    \centering
    \includegraphics[width=\linewidth]{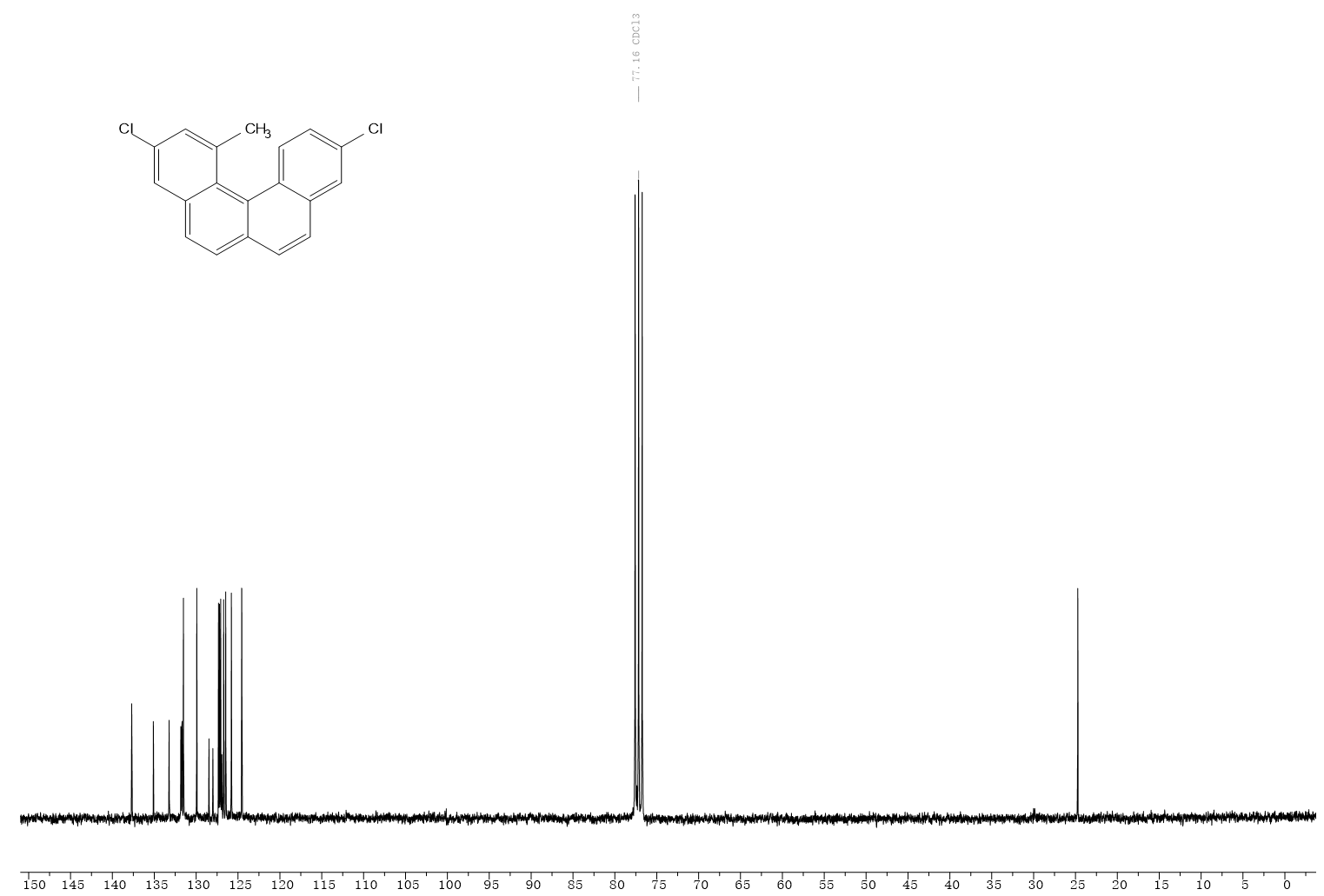}
    \caption{$^{13}$C\,NMR (76\,MHz, CDCl$_3$) of \compound{7}.}
    \label{fig:15}
\end{figure}

\subsection{Synthesis of Compound \compound{8}}
\begin{figure}[htbp]
    \centering
    \includegraphics[width=\linewidth]{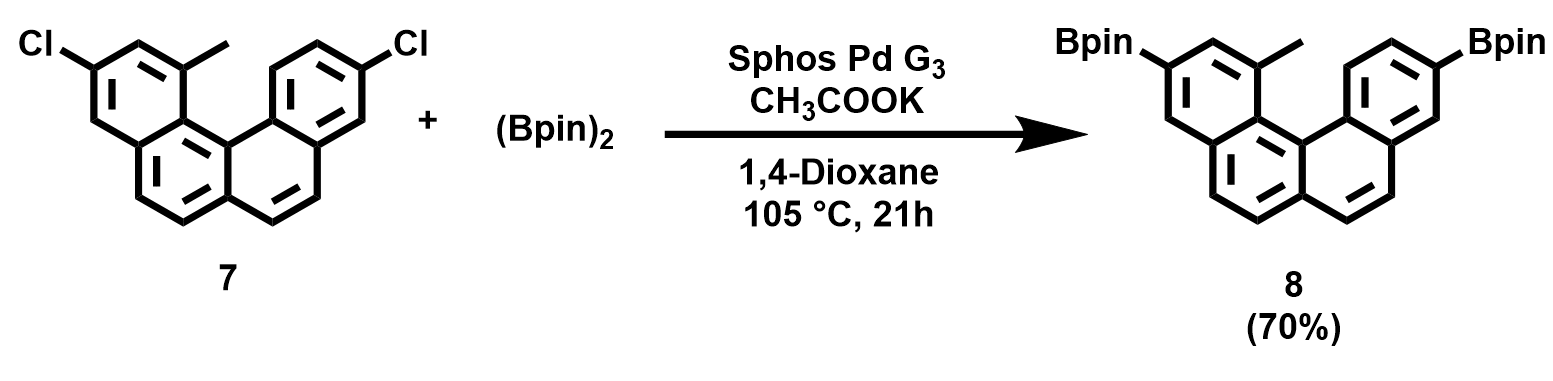}
    \label{fig:c8}
\end{figure} 

Compound \compound{7} (19.8\,mg, 64\,\textmu mol), bis(pinacolato)diboron
(230\,mg, 0.91\,mmol), potassium acetate (98\,mg, 1.00\,mmol), and SPhos Pd G3
(37\,mg, 0.05\,mmol) were weighed into a Schlenk flask under an argon atmosphere.
The flask was purged three times. Degassed 1,4-dioxane (5\,mL) was added via
syringe, and the resulting suspension was stirred at 105\,\textdegree C for 21\,h
under argon. After cooling to room temperature, the mixture was diluted with
dichloromethane, washed with brine, dried over anhydrous magnesium sulfate,
filtered, and concentrated under reduced pressure. The crude product was purified
by silica gel column chromatography (eluent: iso-hexane/DCM\,=\,$1:1 \to 1:2$,
then \ch{CH2Cl2}/ethyl acetate\,=\,$5:1 \to 5:2$) to afford compound \compound{8}
(22\,mg, 70\%) as a brown solid.

\medskip
\noindent
$^{1}$H\,NMR (300\,MHz, \nmrsolv{CDCl$_3$}): $\delta$\,8.49 (d, $J = 1.3$\,Hz,
1H), 8.33 (d, $J = 1.2$\,Hz, 1H), 8.02--7.93 (m, 3H), 7.93--7.90 (m, 1H),
7.90--7.82 (m, 2H), 7.73 (d, $J = 8.4$\,Hz, 1H), 2.41 (s, 3H), 1.43 (dd,
$J = 4.2$, $2.6$\,Hz, 24H).

\medskip
\noindent
$^{13}$C\,NMR (76\,MHz, \nmrsolv{CDCl$_3$}): $\delta$\,135.49, 134.70, 133.22,
130.08, 129.27, 128.69, 128.39, 126.04, 125.86, 84.15, 84.11, 83.63, 25.15, 25.10.

\medskip
\noindent
HR-MS MALDI-TOF ($m/z$): calculated for \ch{C31H36B2O4} [M$+$H]$^+$, 492.2872;
found, 492.2863; error $= -1.88$\,ppm.
\medskip
\noindent
\begin{figure}[htbp]
    \centering
    \includegraphics[width=\linewidth]{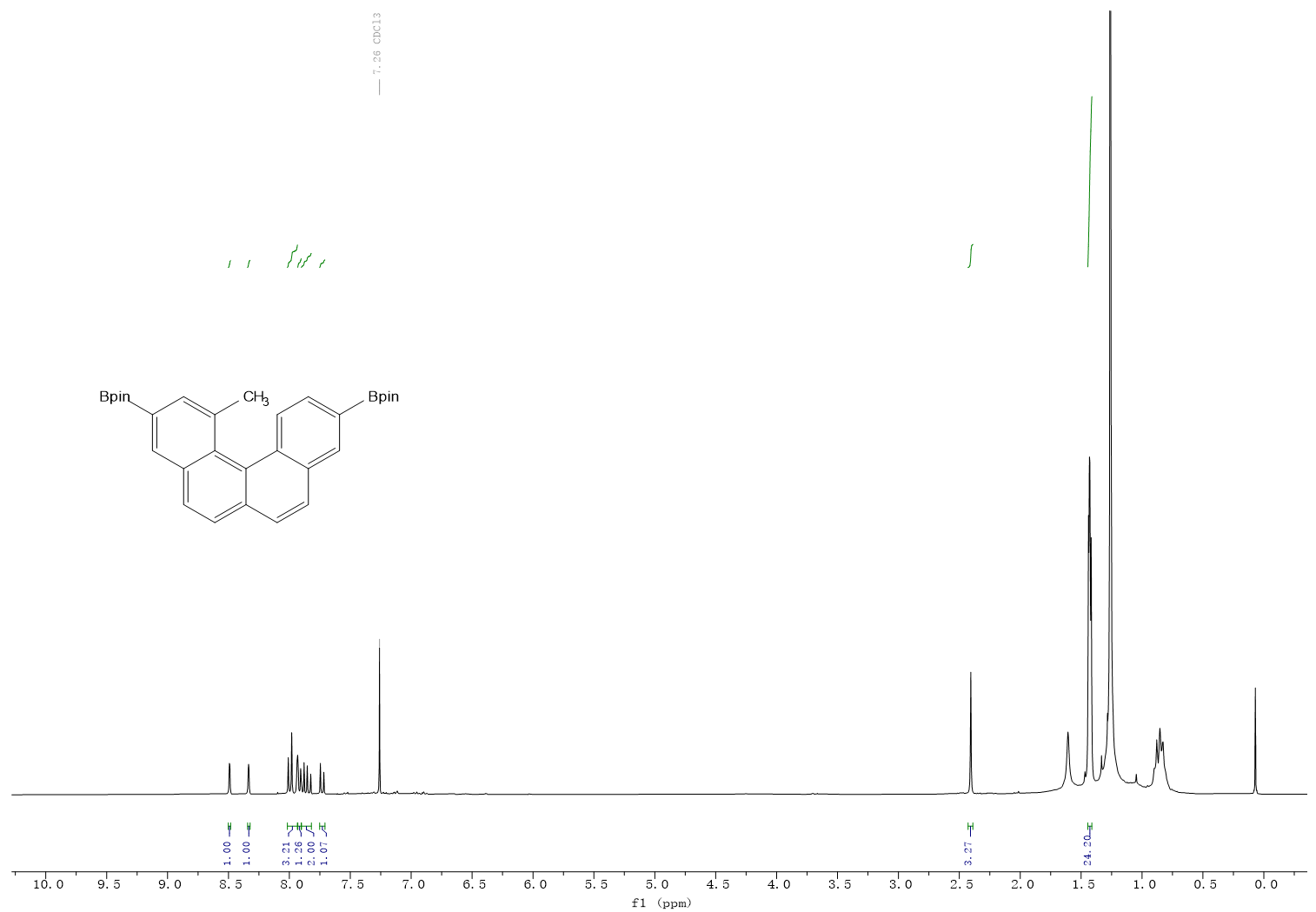}
    \caption{$^{1}$H\,NMR (300\,MHz, CDCl$_3$) of \compound{8}}
    \label{fig:16}
\end{figure}
\medskip
\noindent
\begin{figure}[htbp]
    \centering
    \includegraphics[width=\linewidth]{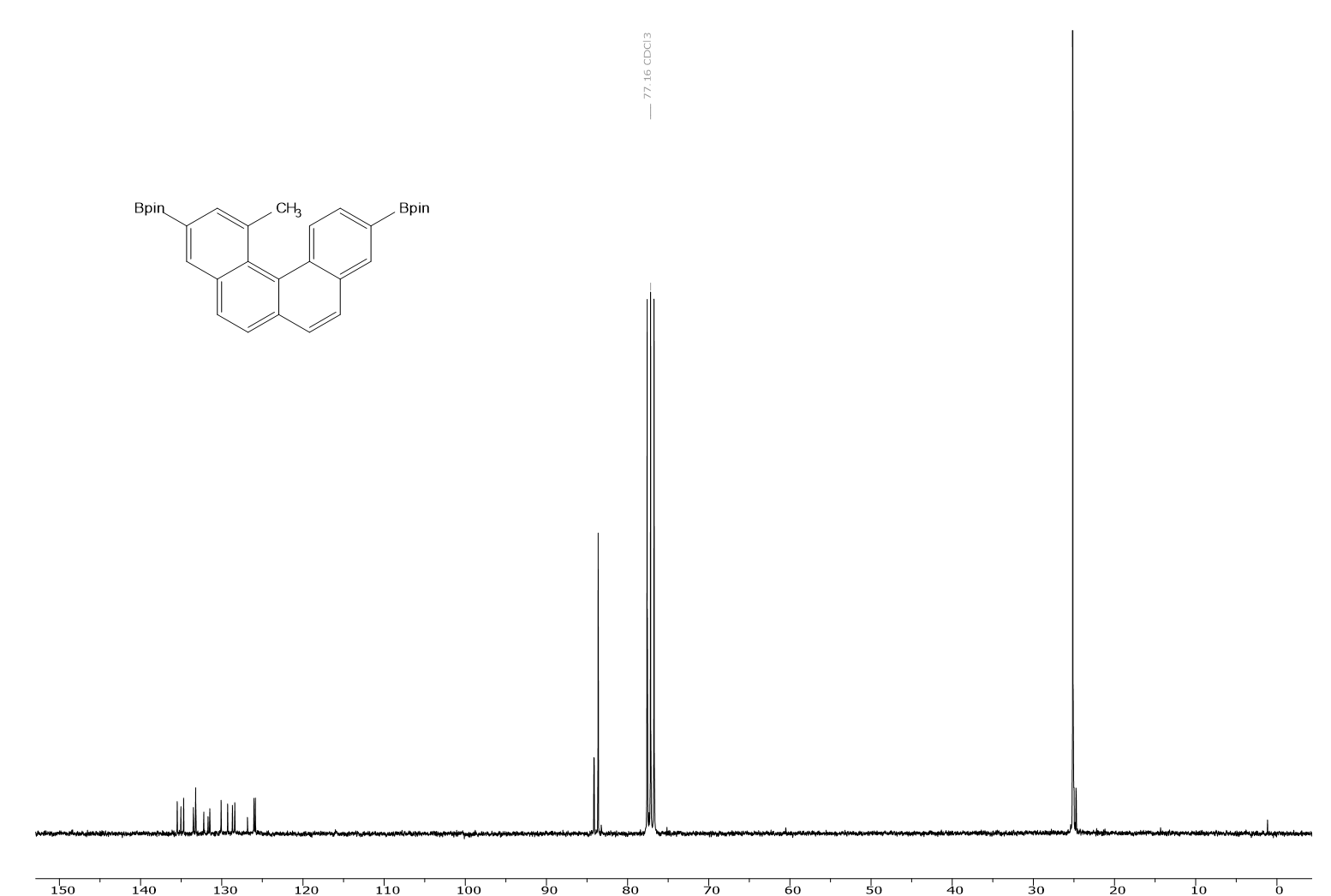}
    \caption{$^{13}$C\,NMR (76\,MHz, CDCl$_3$) of \compound{8}.}
    \label{fig:17}
\end{figure}

\subsection{Synthesis of Compound \compound{9}}
\begin{figure}[htbp]
    \centering
    \includegraphics[width=\linewidth]{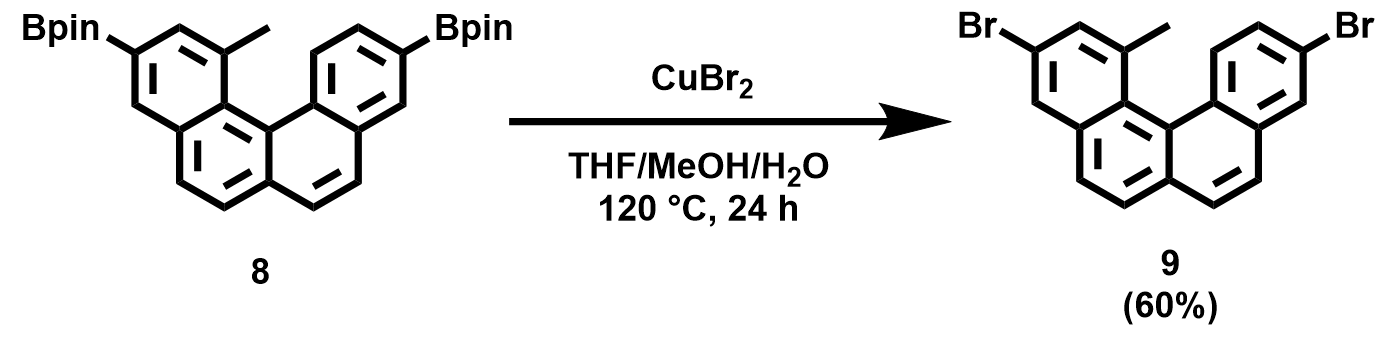}
    \label{fig:c9}
\end{figure} 

Compound \compound{8} (35\,mg, 0.088\,mmol) and \ch{CuBr2} (206\,mg,
0.92\,mmol) were added to a 10\,mL pressure tube. The tube was evacuated and
backfilled with argon three times. A mixed solvent of
\ch{THF}/\ch{MeOH}/\ch{H2O} (1\,mL/4\,mL/2\,mL, previously degassed) was added
via syringe, and the tube was sealed tightly. The mixture was heated to
120\,\textdegree C and stirred for 24\,h. After cooling to room temperature, the
solvent was removed under reduced pressure. The mixture was diluted with
dichloromethane, washed with water, dried over anhydrous magnesium sulfate,
filtered, and concentrated under reduced pressure. The crude product was purified
by silica gel column chromatography (eluent: iso-hexane/DCM\,=\,$1:0 \to 4:1$)
to afford the desired product \compound{9} (17\,mg, 60\%) as a white solid.

\medskip
\noindent
$^{1}$H\,NMR (300\,MHz, \nmrsolv{CDCl$_3$}): $\delta$\,8.14 (d, $J = 2.1$\,Hz,
1H), 8.00--7.98 (m, 1H), 7.86 (s, 2H), 7.82 (d, $J = 9.0$\,Hz, 1H), 7.75
(s, 2H), 7.66--7.61 (m, 2H), 2.38 (s, 3H).

\medskip
\noindent
$^{13}$C\,NMR (76\,MHz, \nmrsolv{CDCl$_3$}): $\delta$\,137.83, 135.44, 133.63,
132.54, 131.83, 131.66, 129.72, 128.72, 128.39, 128.25, 127.74, 127.32, 127.18,
127.07, 126.98, 126.72, 120.12, 119.92, 24.68.

\medskip
\noindent
HR-MS MALDI-TOF ($m/z$): calculated for \ch{C19H12Br2} [M]$^+$, 397.9306;
found, 397.9306; error $= +0.03$\,ppm.
\medskip
\noindent
\begin{figure}[htbp]
    \centering
    \includegraphics[width=\linewidth]{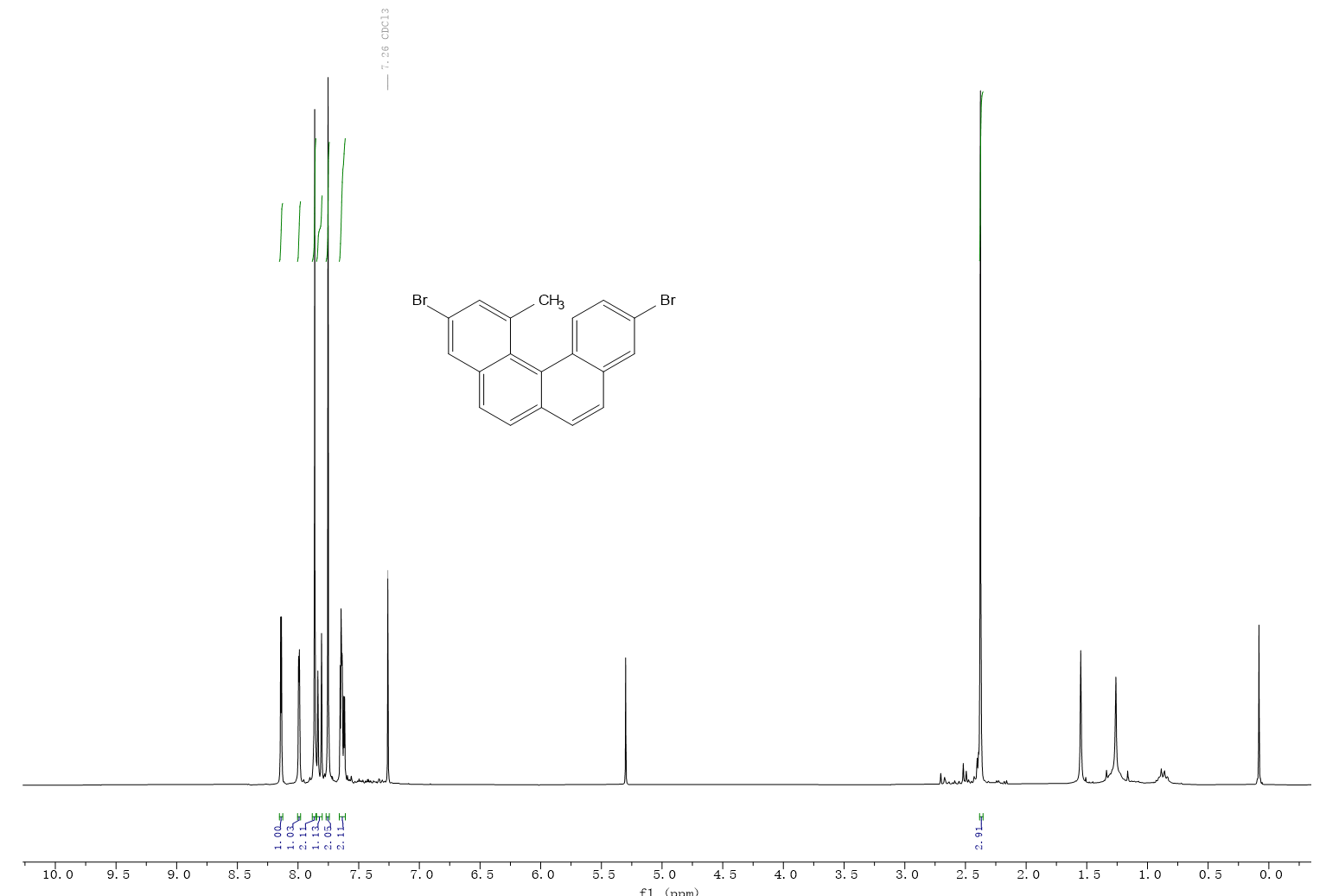}
    \caption{$^{1}$H\,NMR (300\,MHz, CDCl$_3$) of \compound{9}.}
    \label{fig:18}
\end{figure}
\medskip
\noindent
\begin{figure}[htbp]
    \centering
    \includegraphics[width=\linewidth]{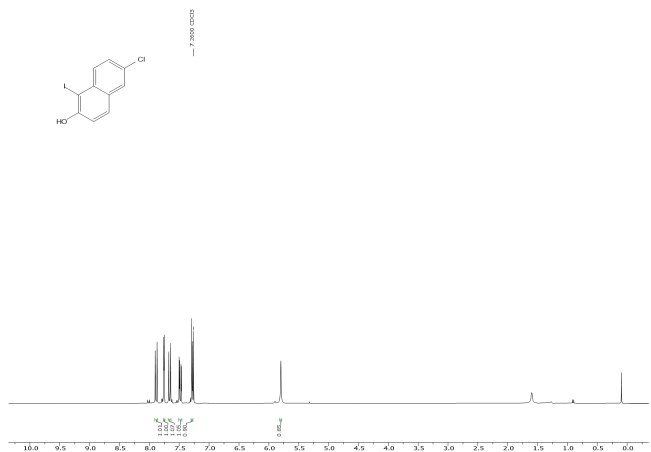}
    \caption{$^{13}$C\,NMR (76\,MHz, CDCl$_3$) of \compound{9}.}
    \label{fig:19}
\end{figure}

\end{document}